\documentclass[thmsa,11pt,a4paper]{article}
\usepackage{amssymb}

\usepackage{sw20lart}

\makeatletter
\newcount\@hour\newcount\@minute\chardef\@x10\chardef\@xv60
\def\tcitime{
\def\@time{%
  \@minute\time\@hour\@minute\divide\@hour\@xv
  \ifnum\@hour<\@x 0\fi\the\@hour:%
  \multiply\@hour\@xv\advance\@minute-\@hour
  \ifnum\@minute<\@x 0\fi\the\@minute
  }}%

\@ifundefined{hyperref}{}{}

\@ifundefined{qExtProgCall}{\def\qExtProgCall#1#2#3#4#5#6{\relax}}{}
\def\QCTOpt[#1]#2{%
  \def\QCTOptB{#1}
  \def\QCTOptA{#2}
}
\def\QCTNOpt#1{%
  \def\QCTOptA{#1}
  \let\QCTOptB\empty
}
\def\Qct{%
  \@ifnextchar[{%
    \QCTOpt}{\QCTNOpt}
}
\def\QCBOpt[#1]#2{%
  \def\QCBOptB{#1}
  \def\QCBOptA{#2}
}
\def\QCBNOpt#1{%
  \def\QCBOptA{#1}
  \let\QCBOptB\empty
}
\def\Qcb{%
  \@ifnextchar[{%
    \QCBOpt}{\QCBNOpt}
}
\def\PrepCapArgs{%
  \ifx\QCBOptA\empty
    \ifx\QCTOptA\empty
      {}%
    \else
      \ifx\QCTOptB\empty
        {\QCTOptA}%
      \else
        [\QCTOptB]{\QCTOptA}%
      \fi
    \fi
  \else
    \ifx\QCBOptA\empty
      {}%
    \else
      \ifx\QCBOptB\empty
        {\QCBOptA}%
      \else
        [\QCBOptB]{\QCBOptA}%
      \fi
    \fi
  \fi
}
\newcount\GRAPHICSTYPE
\GRAPHICSTYPE=\z@
\def\GRAPHICSPS#1{%
 \ifcase\GRAPHICSTYPE
   \special{ps: #1}%
 \or
   \special{language "PS", include "#1"}%
 \fi
}%
\def\graffile#1#2#3#4{%
    \leavevmode
    \raise -#4 \BOXTHEFRAME{%
        \hbox to #2{\raise #3\hbox to #2{\null #1\hfil}}}%
}%
\def\draftbox#1#2#3#4{%
 \leavevmode\raise -#4 \hbox{%
  \frame{\rlap{\protect\tiny #1}\hbox to #2%
   {\vrule height#3 width\z@ depth\z@\hfil}%
  }%
 }%
}%
\newcount\draft
\draft=\z@

\newif\ifwasdraft
\wasdraftfalse

\def\GRAPHIC#1#2#3#4#5{%
 \ifnum\draft=\@ne\draftbox{#2}{#3}{#4}{#5}%
  \else\graffile{#1}{#3}{#4}{#5}%
  \fi
 }%
\def\addtoLaTeXparams#1{%
    \edef\LaTeXparams{\LaTeXparams #1}}%
\newif\ifBoxFrame \BoxFramefalse
\newif\ifOverFrame \OverFramefalse
\newif\ifUnderFrame \UnderFramefalse

\def\BOXTHEFRAME#1{%
   \hbox{%
      \ifBoxFrame
         \frame{#1}%
      \else
         {#1}%
      \fi
   }%
}

\def\doFRAMEparams#1{\BoxFramefalse\OverFramefalse\UnderFramefalse\readFRAMEparams#1\end}%
\def\readFRAMEparams#1{%
 \ifx#1\end%
  \let\next=\relax
  \else
  \ifx#1i\dispkind=\z@\fi
  \ifx#1d\dispkind=\@ne\fi
  \ifx#1f\dispkind=\tw@\fi
  \ifx#1t\addtoLaTeXparams{t}\fi
  \ifx#1b\addtoLaTeXparams{b}\fi
  \ifx#1p\addtoLaTeXparams{p}\fi
  \ifx#1h\addtoLaTeXparams{h}\fi
  \ifx#1X\BoxFrametrue\fi
  \ifx#1O\OverFrametrue\fi
  \ifx#1U\UnderFrametrue\fi
  \ifx#1w
    \ifnum\draft=1\wasdrafttrue\else\wasdraftfalse\fi
    \draft=\@ne
  \fi
  \let\next=\readFRAMEparams
  \fi
 \next
 }%
\def\IFRAME#1#2#3#4#5#6{%
      \bgroup
      \let\QCTOptA\empty
      \let\QCTOptB\empty
      \let\QCBOptA\empty
      \let\QCBOptB\empty
      #6%
      \parindent=0pt%
      \leftskip=0pt
      \rightskip=0pt
      \setbox0 = \hbox{\QCBOptA}%
      \@tempdima = #1\relax
      \ifOverFrame
          \typeout{This is not implemented yet}%
          \show\HELP
      \else
         \ifdim\wd0>\@tempdima
            \advance\@tempdima by \@tempdima
            \ifdim\wd0 >\@tempdima
               \textwidth=\@tempdima
               \setbox1 =\vbox{%
                  \noindent\hbox to \@tempdima{\hfill\GRAPHIC{#5}{#4}{#1}{#2}{#3}\hfill}\\%
                  \noindent\hbox to \@tempdima{\parbox[b]{\@tempdima}{\QCBOptA}}%
               }%
               \wd1=\@tempdima
            \else
               \textwidth=\wd0
               \setbox1 =\vbox{%
                 \noindent\hbox to \wd0{\hfill\GRAPHIC{#5}{#4}{#1}{#2}{#3}\hfill}\\%
                 \noindent\hbox{\QCBOptA}%
               }%
               \wd1=\wd0
            \fi
         \else
            \ifdim\wd0>0pt
              \hsize=\@tempdima
              \setbox1 =\vbox{%
                \unskip\GRAPHIC{#5}{#4}{#1}{#2}{0pt}%
                \break
                \unskip\hbox to \@tempdima{\hfill \QCBOptA\hfill}%
              }%
              \wd1=\@tempdima
           \else
              \hsize=\@tempdima
              \setbox1 =\vbox{%
                \unskip\GRAPHIC{#5}{#4}{#1}{#2}{0pt}%
              }%
              \wd1=\@tempdima
           \fi
         \fi
         \@tempdimb=\ht1
         \advance\@tempdimb by \dp1
         \advance\@tempdimb by -#2%
         \advance\@tempdimb by #3%
         \leavevmode
         \raise -\@tempdimb \hbox{\box1}%
      \fi
      \egroup%
}%
\def\DFRAME#1#2#3#4#5{%
 \begin{center}
     \let\QCTOptA\empty
     \let\QCTOptB\empty
     \let\QCBOptA\empty
     \let\QCBOptB\empty
     \ifOverFrame 
        #5\QCTOptA\par
     \fi
     \GRAPHIC{#4}{#3}{#1}{#2}{\z@}
     \ifUnderFrame 
        \nobreak\par #5\QCBOptA
     \fi
 \end{center}%
 }%
\def\FFRAME#1#2#3#4#5#6#7{%
 \begin{figure}[#1]%
  \let\QCTOptA\empty
  \let\QCTOptB\empty
  \let\QCBOptA\empty
  \let\QCBOptB\empty
  \ifOverFrame
    #4
    \ifx\QCTOptA\empty
    \else
      \ifx\QCTOptB\empty
        \caption{\QCTOptA}%
      \else
        \caption[\QCTOptB]{\QCTOptA}%
      \fi
    \fi
    \ifUnderFrame\else
      \label{#5}%
    \fi
  \else
    \UnderFrametrue%
  \fi
  \begin{center}\GRAPHIC{#7}{#6}{#2}{#3}{\z@}\end{center}%
  \ifUnderFrame
    #4
    \ifx\QCBOptA\empty
      \caption{}%
    \else
      \ifx\QCBOptB\empty
        \caption{\QCBOptA}%
      \else
        \caption[\QCBOptB]{\QCBOptA}%
      \fi
    \fi
    \label{#5}%
  \fi
  \end{figure}%
 }%
\newcount\dispkind%

\def\makeactives{
  \catcode`\"=\active
  \catcode`\;=\active
  \catcode`\:=\active
  \catcode`\'=\active
  \catcode`\~=\active
}
\bgroup
   \makeactives
   \gdef\activesoff{%
      \def"{\string"}
      \def;{\string;}
      \def:{\string:}
      \def'{\string'}
      \def~{\string~}
    }
\egroup

\def\FRAME#1#2#3#4#5#6#7#8{%
 \bgroup
 \@ifundefined{bbl@deactivate}{}{\activesoff}
 \ifnum\draft=\@ne
   \wasdrafttrue
 \else
   \wasdraftfalse%
 \fi
 \def\LaTeXparams{}%
 \dispkind=\z@
 \def\LaTeXparams{}%
 \doFRAMEparams{#1}%
 \ifnum\dispkind=\z@\IFRAME{#2}{#3}{#4}{#7}{#8}{#5}\else
  \ifnum\dispkind=\@ne\DFRAME{#2}{#3}{#7}{#8}{#5}\else
   \ifnum\dispkind=\tw@
    \edef\@tempa{\noexpand\FFRAME{\LaTeXparams}}%
    \@tempa{#2}{#3}{#5}{#6}{#7}{#8}%
    \fi
   \fi
  \fi
  \ifwasdraft\draft=1\else\draft=0\fi{}%
  \egroup
 }%
\def\TEXUX#1{"texux"}

\def\func#1{\mathop{\rm #1}}%
\long\def\QQQ#1#2{%
     \long\expandafter\def\csname#1\endcsname{#2}}%
\@ifundefined{QTP}{\def\QTP#1{}}{}
\@ifundefined{QEXCLUDE}{\def\QEXCLUDE#1{}}{}
\@ifundefined{Qlb}{}{}
\@ifundefined{Qlt}{}{}
\long\def\QQA#1#2{}%
\def\QTR#1#2{{\csname#1\endcsname #2}}
\def\EXPAND#1[#2]#3{}%
\def\NOEXPAND#1[#2]#3{}%
\def\LaTeXparent#1{}%
\def\ChildStyles#1{}%
\def\ChildDefaults#1{}%
\def\QTagDef#1#2#3{}%
\@ifundefined{StyleEditBeginDoc}{}{}
\def\QQfnmark#1{\footnotemark}

\@ifundefined{INDEX}{\def\INDEX#1#2{}{}}{}%
\@ifundefined{SUBINDEX}{\def\SUBINDEX#1#2#3{}{}{}}{}%
\@ifundefined{initial}%
   {\def\initial#1{\bigbreak{\raggedright\large\bf #1}\kern 2\p@\penalty3000}}%
   {}%
\@ifundefined{entry}{}{}%
\@ifundefined{primary}{}{}%
\@ifundefined{secondary}{}{}%
\@ifundefined{ZZZ}{}{\makeatletter\input gnuindex.sty\makeatother\makeindex\makeatletter}%
\@ifundefined{abstract}{%
 \def\abstract{%
  \if@twocolumn
   \section*{Abstract (Not appropriate in this style!)}%
   \else \small 
   \begin{center}{\bf Abstract\vspace{-.5em}\vspace{\z@}}\end{center}%
   \quotation 
   \fi
  }%
 }{%
 }%
\@ifundefined{endabstract}{\def\endabstract
  {\if@twocolumn\else\endquotation\fi}}{}%
\@ifundefined{maketitle}{\def\maketitle#1{}}{}%
\@ifundefined{affiliation}{\def\affiliation#1{}}{}%
\@ifundefined{proof}{}{}%
\@ifundefined{endproof}{}{}%
\@ifundefined{newfield}{\def\newfield#1#2{}}{}%
\@ifundefined{chapter}{\def\chapter#1{\par(Chapter head:)#1\par }%
 \newcount\c@chapter}{}%
\@ifundefined{part}{\def\part#1{\par(Part head:)#1\par }}{}%
\@ifundefined{section}{\def\section#1{\par(Section head:)#1\par }}{}%
\@ifundefined{subsection}{\def\subsection#1%
 {\par(Subsection head:)#1\par }}{}%
\@ifundefined{subsubsection}{\def\subsubsection#1%
 {\par(Subsubsection head:)#1\par }}{}%
\@ifundefined{paragraph}{\def\paragraph#1%
 {\par(Subsubsubsection head:)#1\par }}{}%
\@ifundefined{subparagraph}{\def\subparagraph#1%
 {\par(Subsubsubsubsection head:)#1\par }}{}%
\@ifundefined{therefore}{}{}%
\@ifundefined{backepsilon}{}{}%
\@ifundefined{yen}{}{}%
\@ifundefined{registered}{%
   \def\registered{\relax\ifmmode{}\r@gistered
                    \else$\m@th\r@gistered$\fi}%
 \def\r@gistered{^{\ooalign
  {\hfil\raise.07ex\hbox{$\scriptstyle\rm\text{R}$}\hfil\crcr
  \mathhexbox20D}}}}{}%
\@ifundefined{Eth}{}{}%
\@ifundefined{eth}{}{}%
\@ifundefined{Thorn}{}{}%
\@ifundefined{thorn}{}{}%
\@ifundefined{degree}{}{}%
\newdimen\theight
\def\Column{%
 \vadjust{\setbox\z@=\hbox{\scriptsize\quad\quad tcol}%
  \theight=\ht\z@\advance\theight by \dp\z@\advance\theight by \lineskip
  \kern -\theight \vbox to \theight{%
   \rightline{\rlap{\box\z@}}%
   \vss
   }%
  }%
 }%
\def\qed{%
 \ifhmode\unskip\nobreak\fi\ifmmode\ifinner\else\hskip5\p@\fi\fi
 \hbox{\hskip5\p@\vrule width4\p@ height6\p@ depth1.5\p@\hskip\p@}%
 }%
\def\miss{\hbox{\vrule height2\p@ width 2\p@ depth\z@}}%
\def\tcol#1{{\baselineskip=6\p@ \vcenter{#1}} \Column}  %
\def\newfmtname{LaTeX2e}
\def\chkcompat{%
   \if@compatibility
   \else
     \usepackage{latexsym}
   \fi
}

\ifx\fmtname\newfmtname
  \DeclareOldFontCommand{\rm}{\normalfont\rmfamily}{\mathrm}
  \DeclareOldFontCommand{\sf}{\normalfont\sffamily}{\mathsf}
  \DeclareOldFontCommand{\tt}{\normalfont\ttfamily}{\mathtt}
  \DeclareOldFontCommand{\bf}{\normalfont\bfseries}{\mathbf}
  \DeclareOldFontCommand{\it}{\normalfont\itshape}{\mathit}
  \DeclareOldFontCommand{\sl}{\normalfont\slshape}{\@nomath\sl}
  \DeclareOldFontCommand{\sc}{\normalfont\scshape}{\@nomath\sc}
  \chkcompat
\fi

\def\alpha{{\Greekmath 010B}}%
\def\beta{{\Greekmath 010C}}%
\def\gamma{{\Greekmath 010D}}%
\def\delta{{\Greekmath 010E}}%
\def\epsilon{{\Greekmath 010F}}%
\def\zeta{{\Greekmath 0110}}%
\def\eta{{\Greekmath 0111}}%
\def\theta{{\Greekmath 0112}}%
\def\iota{{\Greekmath 0113}}%
\def\kappa{{\Greekmath 0114}}%
\def\lambda{{\Greekmath 0115}}%
\def\mu{{\Greekmath 0116}}%
\def\nu{{\Greekmath 0117}}%
\def\xi{{\Greekmath 0118}}%
\def\pi{{\Greekmath 0119}}%
\def\rho{{\Greekmath 011A}}%
\def\sigma{{\Greekmath 011B}}%
\def\tau{{\Greekmath 011C}}%
\def\upsilon{{\Greekmath 011D}}%
\def\phi{{\Greekmath 011E}}%
\def\chi{{\Greekmath 011F}}%
\def\psi{{\Greekmath 0120}}%
\def\omega{{\Greekmath 0121}}%
\def\varepsilon{{\Greekmath 0122}}%
\def\vartheta{{\Greekmath 0123}}%
\def\varpi{{\Greekmath 0124}}%
\def\varrho{{\Greekmath 0125}}%
\def\varsigma{{\Greekmath 0126}}%
\def\varphi{{\Greekmath 0127}}%

\def\nabla{{\Greekmath 0272}}
\def\FindBoldGroup{%
   {\setbox0=\hbox{$\mathbf{x\global\edef\theboldgroup{\the\mathgroup}}$}}%
}

\def\Greekmath#1#2#3#4{%
    \if@compatibility
        \ifnum\mathgroup=\symbold
           \mathchoice{\mbox{\boldmath$\displaystyle\mathchar"#1#2#3#4$}}%
                      {\mbox{\boldmath$\textstyle\mathchar"#1#2#3#4$}}%
                      {\mbox{\boldmath$\scriptstyle\mathchar"#1#2#3#4$}}%
                      {\mbox{\boldmath$\scriptscriptstyle\mathchar"#1#2#3#4$}}%
        \else
           \mathchar"#1#2#3#4%
        \fi 
    \else 
        \FindBoldGroup
        \ifnum\mathgroup=\theboldgroup 
           \mathchoice{\mbox{\boldmath$\displaystyle\mathchar"#1#2#3#4$}}%
                      {\mbox{\boldmath$\textstyle\mathchar"#1#2#3#4$}}%
                      {\mbox{\boldmath$\scriptstyle\mathchar"#1#2#3#4$}}%
                      {\mbox{\boldmath$\scriptscriptstyle\mathchar"#1#2#3#4$}}%
        \else
           \mathchar"#1#2#3#4%
        \fi     	    
	  \fi}

\newif\ifGreekBold  \GreekBoldfalse
\let\SAVEPBF=\pbf
\def\pbf{\GreekBoldtrue\SAVEPBF}%

\@ifundefined{theorem}{}{}
\@ifundefined{lemma}{}{}
\@ifundefined{corollary}{}{}
\@ifundefined{conjecture}{}{}
\@ifundefined{proposition}{}{}
\@ifundefined{axiom}{}{}
\@ifundefined{remark}{}{}
\@ifundefined{example}{}{}
\@ifundefined{exercise}{}{}
\@ifundefined{definition}{}{}

\@ifundefined{mathletters}{%
  \newcounter{equationnumber}  
  \def\mathletters{%
     \addtocounter{equation}{1}
     \edef\@currentlabel{\theequation}%
     \setcounter{equationnumber}{\c@equation}
     \setcounter{equation}{0}%
     \edef\theequation{\@currentlabel\noexpand\alph{equation}}%
  }
  
}{}

\@ifundefined{BibTeX}{%
    \def\BibTeX{{\rm B\kern-.05em{\sc i\kern-.025em b}\kern-.08em
                 T\kern-.1667em\lower.7ex\hbox{E}\kern-.125emX}}}{}%
\@ifundefined{AmS}%
    {\def\AmS{{\protect\usefont{OMS}{cmsy}{m}{n}%
                A\kern-.1667em\lower.5ex\hbox{M}\kern-.125emS}}}{}%
\@ifundefined{AmSTeX}{}{}%
\ifx\ds@amstex\relax
\makeatother\endinput
\else
   \@ifpackageloaded{amstex}%
      {\message{amstex already loaded}\makeatother\endinput}
      {}
   \@ifpackageloaded{amsgen}%
      {\message{amsgen already loaded}\makeatother\endinput}
      {}
\fi
\let\DOTSI\relax
\def\RIfM@{\relax\ifmmode}%
\def\FN@{\futurelet\next}%
\newcount\intno@
\def\iint{\DOTSI\intno@\tw@\FN@\ints@}%
\def\iiint{\DOTSI\intno@\thr@@\FN@\ints@}%
\def\iiiint{\DOTSI\intno@4 \FN@\ints@}%
\def\idotsint{\DOTSI\intno@\z@\FN@\ints@}%
\def\ints@{\findlimits@\ints@@}%
\newif\iflimtoken@
\newif\iflimits@
\def\findlimits@{\limtoken@true\ifx\next\limits\limits@true
 \else\ifx\next\nolimits\limits@false\else
 \limtoken@false\ifx\ilimits@\nolimits\limits@false\else
 \ifinner\limits@false\else\limits@true\fi\fi\fi\fi}%
\def\multint@{\int\ifnum\intno@=\z@\intdots@                          
 \else\intkern@\fi                                                    
 \ifnum\intno@>\tw@\int\intkern@\fi                                   
 \ifnum\intno@>\thr@@\int\intkern@\fi                                 
 \int}
\def\multintlimits@{\intop\ifnum\intno@=\z@\intdots@\else\intkern@\fi
 \ifnum\intno@>\tw@\intop\intkern@\fi
 \ifnum\intno@>\thr@@\intop\intkern@\fi\intop}%
\def\intic@{%
    \mathchoice{\hskip.5em}{\hskip.4em}{\hskip.4em}{\hskip.4em}}%
\def\negintic@{\mathchoice
 {\hskip-.5em}{\hskip-.4em}{\hskip-.4em}{\hskip-.4em}}%
\def\ints@@{\iflimtoken@                                              
 \def\ints@@@{\iflimits@\negintic@
   \mathop{\intic@\multintlimits@}\limits                             
  \else\multint@\nolimits\fi                                          
  \eat@}
 \else                                                                
 \def\ints@@@{\iflimits@\negintic@
  \mathop{\intic@\multintlimits@}\limits\else
  \multint@\nolimits\fi}\fi\ints@@@}%
\def\intkern@{\mathchoice{\!\!\!}{\!\!}{\!\!}{\!\!}}%
\def\plaincdots@{\mathinner{\cdotp\cdotp\cdotp}}%
\def\intdots@{\mathchoice{\plaincdots@}%
 {{\cdotp}\mkern1.5mu{\cdotp}\mkern1.5mu{\cdotp}}%
 {{\cdotp}\mkern1mu{\cdotp}\mkern1mu{\cdotp}}%
 {{\cdotp}\mkern1mu{\cdotp}\mkern1mu{\cdotp}}}%
\def\RIfM@{\relax\protect\ifmmode}
\def\text{\RIfM@\expandafter\text@\else\expandafter\mbox\fi}
\let\nfss@text\text
\def\text@#1{\mathchoice
   {\textdef@\displaystyle\f@size{#1}}%
   {\textdef@\textstyle\tf@size{\firstchoice@false #1}}%
   {\textdef@\textstyle\sf@size{\firstchoice@false #1}}%
   {\textdef@\textstyle \ssf@size{\firstchoice@false #1}}%
   \glb@settings}

\def\textdef@#1#2#3{\hbox{{%
                    \everymath{#1}%
                    \let\f@size#2\selectfont
                    #3}}}
\newif\iffirstchoice@
\firstchoice@true
\def\Let@{\relax\iffalse{\fi\let\\=\cr\iffalse}\fi}%
\def\vspace@{\def\vspace##1{\crcr\noalign{\vskip##1\relax}}}%
\def\multilimits@{\bgroup\vspace@\Let@
 \baselineskip\fontdimen10 \scriptfont\tw@
 \advance\baselineskip\fontdimen12 \scriptfont\tw@
 \lineskip\thr@@\fontdimen8 \scriptfont\thr@@
 \lineskiplimit\lineskip
 \vbox\bgroup\ialign\bgroup\hfil$\m@th\scriptstyle{##}$\hfil\crcr}%
\def\Sb{_\multilimits@}%
\def\endSb{\crcr\egroup\egroup\egroup}%
\def\Sp{^\multilimits@}%

\newdimen\ex@
\def\rightarrowfill@#1{$#1\m@th\mathord-\mkern-6mu\cleaders
 \hbox{$#1\mkern-2mu\mathord-\mkern-2mu$}\hfill
 \mkern-6mu\mathord\rightarrow$}%
\def\leftarrowfill@#1{$#1\m@th\mathord\leftarrow\mkern-6mu\cleaders
 \hbox{$#1\mkern-2mu\mathord-\mkern-2mu$}\hfill\mkern-6mu\mathord-$}%
\def\leftrightarrowfill@#1{$#1\m@th\mathord\leftarrow
\mkern-6mu\cleaders
 \hbox{$#1\mkern-2mu\mathord-\mkern-2mu$}\hfill
 \mkern-6mu\mathord\rightarrow$}%
\def\overrightarrow{\mathpalette\overrightarrow@}%
\def\overrightarrow@#1#2{\vbox{\ialign{##\crcr\rightarrowfill@#1\crcr
 \noalign{\kern-\ex@\nointerlineskip}$\m@th\hfil#1#2\hfil$\crcr}}}%

\def\overleftarrow{\mathpalette\overleftarrow@}%
\def\overleftarrow@#1#2{\vbox{\ialign{##\crcr\leftarrowfill@#1\crcr
 \noalign{\kern-\ex@\nointerlineskip}$\m@th\hfil#1#2\hfil$\crcr}}}%
\def\overleftrightarrow{\mathpalette\overleftrightarrow@}%
\def\overleftrightarrow@#1#2{\vbox{\ialign{##\crcr
   \leftrightarrowfill@#1\crcr
 \noalign{\kern-\ex@\nointerlineskip}$\m@th\hfil#1#2\hfil$\crcr}}}%
\def\underrightarrow{\mathpalette\underrightarrow@}%
\def\underrightarrow@#1#2{\vtop{\ialign{##\crcr$\m@th\hfil#1#2\hfil
  $\crcr\noalign{\nointerlineskip}\rightarrowfill@#1\crcr}}}%

\def\underleftarrow{\mathpalette\underleftarrow@}%
\def\underleftarrow@#1#2{\vtop{\ialign{##\crcr$\m@th\hfil#1#2\hfil
  $\crcr\noalign{\nointerlineskip}\leftarrowfill@#1\crcr}}}%
\def\underleftrightarrow{\mathpalette\underleftrightarrow@}%
\def\underleftrightarrow@#1#2{\vtop{\ialign{##\crcr$\m@th
  \hfil#1#2\hfil$\crcr
 \noalign{\nointerlineskip}\leftrightarrowfill@#1\crcr}}}%

\def\qopnamewl@#1{\mathop{\operator@font#1}\nlimits@}
\let\nlimits@\displaylimits
\def\setboxz@h{\setbox\z@\hbox}

\def\varlim@#1#2{\mathop{\vtop{\ialign{##\crcr
 \hfil$#1\m@th\operator@font lim$\hfil\crcr
 \noalign{\nointerlineskip}#2#1\crcr
 \noalign{\nointerlineskip\kern-\ex@}\crcr}}}}

 \def\rightarrowfill@#1{\m@th\setboxz@h{$#1-$}\ht\z@\z@
  $#1\copy\z@\mkern-6mu\cleaders
  \hbox{$#1\mkern-2mu\box\z@\mkern-2mu$}\hfill
  \mkern-6mu\mathord\rightarrow$}
\def\leftarrowfill@#1{\m@th\setboxz@h{$#1-$}\ht\z@\z@
  $#1\mathord\leftarrow\mkern-6mu\cleaders
  \hbox{$#1\mkern-2mu\copy\z@\mkern-2mu$}\hfill
  \mkern-6mu\box\z@$}

\def\projlim{\qopnamewl@{proj\,lim}}
\def\injlim{\qopnamewl@{inj\,lim}}
\def\varinjlim{\mathpalette\varlim@\rightarrowfill@}
\def\varprojlim{\mathpalette\varlim@\leftarrowfill@}
\def\varliminf{\mathpalette\varliminf@{}}
\def\varliminf@#1{\mathop{\underline{\vrule\@depth.2\ex@\@width\z@
   \hbox{$#1\m@th\operator@font lim$}}}}
\def\varlimsup{\mathpalette\varlimsup@{}}
\def\varlimsup@#1{\mathop{\overline
  {\hbox{$#1\m@th\operator@font lim$}}}}

\def\tfrac#1#2{{\textstyle {#1 \over #2}}}%
\def\dfrac#1#2{{\displaystyle {#1 \over #2}}}%
\def\dint{\mathop{\displaystyle \int}}%
\begingroup \catcode `|=0 \catcode `[= 1
\catcode`]=2 \catcode `\{=12 \catcode `\}=12
\catcode`\\=12 
|gdef|@alignverbatim#1\end{align}[#1|end[align]]
|gdef|@salignverbatim#1\end{align*}[#1|end[align*]]

|gdef|@alignatverbatim#1\end{alignat}[#1|end[alignat]]
|gdef|@salignatverbatim#1\end{alignat*}[#1|end[alignat*]]

|gdef|@xalignatverbatim#1\end{xalignat}[#1|end[xalignat]]
|gdef|@sxalignatverbatim#1\end{xalignat*}[#1|end[xalignat*]]

|gdef|@gatherverbatim#1\end{gather}[#1|end[gather]]
|gdef|@sgatherverbatim#1\end{gather*}[#1|end[gather*]]

|gdef|@gatherverbatim#1\end{gather}[#1|end[gather]]
|gdef|@sgatherverbatim#1\end{gather*}[#1|end[gather*]]

|gdef|@multilineverbatim#1\end{multiline}[#1|end[multiline]]
|gdef|@smultilineverbatim#1\end{multiline*}[#1|end[multiline*]]

|gdef|@arraxverbatim#1\end{arrax}[#1|end[arrax]]
|gdef|@sarraxverbatim#1\end{arrax*}[#1|end[arrax*]]

|gdef|@tabulaxverbatim#1\end{tabulax}[#1|end[tabulax]]
|gdef|@stabulaxverbatim#1\end{tabulax*}[#1|end[tabulax*]]

|endgroup

\def\align{\@verbatim \frenchspacing\@vobeyspaces \@alignverbatim
You are using the "align" environment in a style in which it is not defined.}

\@namedef{align*}{\@verbatim\@salignverbatim
You are using the "align*" environment in a style in which it is not defined.}
\expandafter\let\csname endalign*\endcsname =\endtrivlist

\def\alignat{\@verbatim \frenchspacing\@vobeyspaces \@alignatverbatim
You are using the "alignat" environment in a style in which it is not defined.}

\@namedef{alignat*}{\@verbatim\@salignatverbatim
You are using the "alignat*" environment in a style in which it is not defined.}
\expandafter\let\csname endalignat*\endcsname =\endtrivlist

\def\xalignat{\@verbatim \frenchspacing\@vobeyspaces \@xalignatverbatim
You are using the "xalignat" environment in a style in which it is not defined.}

\@namedef{xalignat*}{\@verbatim\@sxalignatverbatim
You are using the "xalignat*" environment in a style in which it is not defined.}
\expandafter\let\csname endxalignat*\endcsname =\endtrivlist

\def\gather{\@verbatim \frenchspacing\@vobeyspaces \@gatherverbatim
You are using the "gather" environment in a style in which it is not defined.}

\@namedef{gather*}{\@verbatim\@sgatherverbatim
You are using the "gather*" environment in a style in which it is not defined.}
\expandafter\let\csname endgather*\endcsname =\endtrivlist

\def\multiline{\@verbatim \frenchspacing\@vobeyspaces \@multilineverbatim
You are using the "multiline" environment in a style in which it is not defined.}

\@namedef{multiline*}{\@verbatim\@smultilineverbatim
You are using the "multiline*" environment in a style in which it is not defined.}
\expandafter\let\csname endmultiline*\endcsname =\endtrivlist

\def\arrax{\@verbatim \frenchspacing\@vobeyspaces \@arraxverbatim
You are using a type of "array" construct that is only allowed in AmS-LaTeX.}

\def\tabulax{\@verbatim \frenchspacing\@vobeyspaces \@tabulaxverbatim
You are using a type of "tabular" construct that is only allowed in AmS-LaTeX.}

\@namedef{arrax*}{\@verbatim\@sarraxverbatim
You are using a type of "array*" construct that is only allowed in AmS-LaTeX.}
\expandafter\let\csname endarrax*\endcsname =\endtrivlist

\@namedef{tabulax*}{\@verbatim\@stabulaxverbatim
You are using a type of "tabular*" construct that is only allowed in AmS-LaTeX.}
\expandafter\let\csname endtabulax*\endcsname =\endtrivlist

\def\@@eqncr{\let\@tempa\relax
    \ifcase\@eqcnt \def\@tempa{& & &}\or \def\@tempa{& &}%
      \else \def\@tempa{&}\fi
     \@tempa
     \if@eqnsw
        \iftag@
           \@taggnum
        \else
           \@eqnnum\stepcounter{equation}%
        \fi
     \fi
     \global\tag@false
     \global\@eqnswtrue
     \global\@eqcnt\z@\cr}

 \def\endequation{%
     \ifmmode\ifinner 
      \iftag@
        \addtocounter{equation}{-1} 
        $\hfil
           \displaywidth\linewidth\@taggnum\egroup \endtrivlist
        \global\tag@false
        \global\@ignoretrue   
      \else
        $\hfil
           \displaywidth\linewidth\@eqnnum\egroup \endtrivlist
        \global\tag@false
        \global\@ignoretrue 
      \fi
     \else   
      \iftag@
        \addtocounter{equation}{-1} 
        \eqno \hbox{\@taggnum}
        \global\tag@false%
        $$\global\@ignoretrue
      \else
        \eqno \hbox{\@eqnnum}
        $$\global\@ignoretrue
      \fi
     \fi\fi
 } 

 \newif\iftag@ \tag@false
 
 \def\tag{\@ifnextchar*{\@tagstar}{\@tag}}
 \def\@tag#1{%
     \global\tag@true
     \global\def\@taggnum{(#1)}}
 \def\@tagstar*#1{%
     \global\tag@true
     \global\def\@taggnum{#1}%
}

\makeatother

\begin{document}

\author{G. Lopez Castro$^{1}$, J. Pestieau$^{2}$, C. Smith$^{2}$ and S. Trine$^{2}$. 
\\
\quad \\
\quad \\
\quad \\
$^{1}$\textit{\ Departamento F\'{i}sica, Centro de Investigaci\'{o}n y de
Estudios}\\
\textit{\ Avanzados del IPN, Apdo. Postal 14-740, 07000 M\'{e}xico, D.F.,
M\'{e}xico}\\
$^{2}$\textit{\ Institut de Physique Th\'{e}orique, Universit\'{e}
Catholique de }\\
\textit{Louvain, Chemin du Cyclotron 2, B-1348 Louvain-la-Neuve, Belgium}%
\quad \\
\quad \\
\quad \\
\quad}
\date{}
\title{{\huge Analytical Behaviour of Positronium Decay Amplitudes}\\
\quad \\
\quad \\
\quad \\
}
\maketitle

\begin{abstract}
\quad \newline
\quad \newline
Positronium annihilation amplitudes that are computed by assuming a
factorization approximation with on-shell intermediate leptons do not
exhibit good analytical behaviour.\ Using dispersion techniques, we find new
contributions that interfere with the known results to restore analytical
properties. Those new amplitudes which cannot be obtained using standard
factorized amplitude formalism, contribute at $\mathcal{O}\left( \alpha
^{2}\right) $. Therefore they have to be evaluated before any theoretical
conclusion can be drawn upon the orthopositronium lifetime puzzle.

\quad \newline

\quad \newline

\quad

\quad \newline

\end{abstract}

\section{Introduction}

Positronium is a bound state of electron and positron. In this paper, we
will be interested in the triplet state, orthopositronium, whose decay rate
into $3\gamma $ has been precisely measured : 
\[
\Gamma ^{exp}\left( o\text{-}Ps\rightarrow 3\gamma \right) =\left\{ 
\begin{array}{l}
7.0398\left( 29\right) \,\,\,\mu sec^{-1}\qquad \text{Tokyo\cite{Tokyo}} \\ 
7.0514\left( 14\right) \,\,\,\mu sec^{-1}\qquad \text{Ann Arbor (Gas)\cite
{AnnArborGas}} \\ 
7.0482\left( 16\right) \,\,\,\mu sec^{-1}\qquad \text{Ann Arbor (Vacuum)\cite
{AnnArborVac}}
\end{array}
\right. 
\]

The corresponding theoretical predictions which include perturbative QED
corrections to a non-relativistic treatment of the bound state wavefunction
have been computed also with high accuracy (see for example \cite{OrePowell}%
, \cite{OrthoCorr}, \cite{Adkins}, \cite{OrthoCorr2}): 
\begin{eqnarray*}
\Gamma \left( o\text{-}Ps\rightarrow 3\gamma \right) &=&\alpha ^{6}m\,\frac{%
2\left( \pi ^{2}-9\right) }{9\pi }\left[ 1-A\frac{\alpha }{\pi }-\frac{%
\alpha }{3}^{2}\log \frac{1}{\alpha }+B_{o}\left( \frac{\alpha }{\pi }%
\right) ^{2}-\frac{3\alpha ^{3}}{2\pi }\log ^{2}\frac{1}{\alpha }\right] \\
&=&\left( 7.0382+B_{o}0.39\times 10^{-4}\right) \,\,\,\mu \sec ^{-1}
\end{eqnarray*}
with $A=10.286606\left( 10\right) $, $\alpha $ the fine structure constant
and $m$ the electron mass. Recent theoretical efforts have focused on a more
complete evaluation of the non-logarithmic $\mathcal{O}\left( \alpha
^{2}\right) $ perturbative corrections, with the result $B_{o}=44.52\left(
26\right) $ \cite{Bterm} or, equivalently 
\[
\Gamma \left( o\text{-}Ps\rightarrow 3\gamma \right) =7.039934\left(
10\right) \,\,\,\mu \sec ^{-1} 
\]
This result renders the theoretical prediction still closer to the
experimental measurement of Ref. \cite{Tokyo}.

Positronium is a test ground for bound state treatment in Quantum Field
Theory. The first try dates back to the $40^{\prime }s$, with decay rates
expressed through a factorized formula \cite{History} 
\[
\Gamma \left( o\text{-}Ps\rightarrow 3\gamma \right) =\frac{1}{3}\left| \phi
_{o}\right| ^{2}\cdot \left( 4v_{rel}\sigma \left( e^{+}e^{-}\rightarrow
3\gamma \right) \right) _{v_{rel}\rightarrow 0} 
\]
with $\phi _{o}$ the Schr\"{o}dinger positronium wavefunction at the origin, 
$\sigma \left( e^{+}e^{-}\rightarrow 3\gamma \right) $ the total cross
section for $e^{+}e^{-}\rightarrow 3\gamma $ and $v_{rel}$ the relative
velocity of $e^{+}$ and $e^{-}$ in their center of mass frame. Since then,
more sophisticated decay amplitudes have been constructed, and systematic
procedures for calculating corrections have been developed \cite{NRQED}.
However, the basic factorization of the bound state dynamics from the
annihilation process has remained as a basic postulate. For low order
corrections, this approximation is unquestionable, but for $\mathcal{O}%
\left( \alpha ^{2}\right) $ corrections, factorization has to be tested.
Indeed, non-perturbative phenomena responsible for the off-shellness of the
electron and positron inside the positronium are of $\mathcal{O}\left(
\alpha ^{2}\right) $. In other words, to get a sensible theoretical
prediction at $\mathcal{O}\left( \alpha ^{2}\right) $, one must carefully
analyze how binding energy effects enter the general factorization approach.

In a recent paper \cite{OurWork1}, we showed that the factorization of the
bound state dynamics from the annihilation process can be given a firm
grounding through dispersion relations. From a fully relativistic model, we
recovered the standard factorized amplitudes used in the literature, and
found some forgotten $\mathcal{O}\left( \alpha ^{2}\right) $ contributions.
Those amplitudes are gauge invariant, since they are expressed in terms of
scattering amplitudes for \textit{on-shell} $e^{+}e^{-}$. When applied to
orthopositronium, however, the appearance of those on-shell intermediate
leptons brings infrared singularities \cite{AnsatzWork}, giving an incorrect
analytical behaviour to the amplitude $o$-$Ps\rightarrow 3\gamma $. Indeed,
basic principles require that this amplitude vanishes in the soft photon
limit, since it involves only neutral bosons \cite{Low}. One must conclude
that the standard approaches are not complete and that they can only
approximately describe the positronium decay.

The purpose of the present paper is to restore the analytical behaviour of
positronium decay amplitudes. The relativistic model we used in \cite
{OurWork1} to describe $p$-$Ps\rightarrow \gamma \gamma $ has the correct
analytical behaviour when applied to $o$-$Ps\rightarrow 3\gamma $ since only
off-shell constituents appear. Indeed, we will find along with the known
factorized-type amplitudes already found for parapositronium, a whole class
of non-factorizable processes which enforce the positronium decay amplitude
to vanish in the soft photon limit. Those new amplitudes were absent for $p$-%
$Ps\rightarrow \gamma \gamma $, but will contribute to $o$-$Ps\rightarrow
3\gamma $ at $\mathcal{O}\left( \alpha ^{2}\right) $.

This paper is organized as follows. We first give a brief description of the
derivation of factorized amplitudes from our relativistic model. Then we
analyze the simple decay $p$-$Dm\rightarrow \gamma e^{+}e^{-}$, and show in
detail how the analytical behaviour of the amplitude is restored by new
contributions. Finally, we discuss the orthopositronium decay to three
photons and describe all the new contributions. In the appendix, we present
a short discussion of the decay $K_{S}^{0}\rightarrow \gamma e^{+}e^{-}$,
the hadronic counterpart of the decay $p$-$Dm\rightarrow \gamma e^{+}e^{-}$,
in order to show the generality of the arguments developed for
electromagnetic bound states.

\section{Factorized Amplitude from a Loop Model}

Here we briefly review the recently proposed method \cite{OurWork1} to
derive factorized amplitudes for bound state decays.

The decay rates for positronium are calculated in a loop model. The
positronium first decays into a virtual electron-positron pair, which
subsequently annihilates into real or virtual photons (an odd number for
ortho-states, an even number for para-states). The coupling of the
positronium to its constituents is described by a form factor, denoted by $%
F_{B}$. It is not assumed to be a constant, since a constant form factor
would amount to consider positronium as a point-like state.

In our model, the decay amplitudes are (see for example figure 4) 
\begin{eqnarray}
\mathcal{M}^{\mu \nu ...}\left( ^{1}S\rightarrow A\right) &=&\int \frac{%
d^{4}q}{\left( 2\pi \right) ^{4}}F_{B}Tr\left\{ \gamma _{5}\frac{i}{%
\!\not\!%
q-\frac{1}{2}%
\!\not\!%
P-m}\Gamma ^{\mu \nu ...}\frac{i}{%
\!\not\!%
q+\frac{1}{2}%
\!\not\!%
P-m}\right\}  \label{ParaDec} \\
\mathcal{M}^{\mu \nu ...}\left( ^{3}S\rightarrow A\right) &=&e_{\alpha }\int 
\frac{d^{4}q}{\left( 2\pi \right) ^{4}}F_{B}Tr\left\{ \gamma ^{\alpha }\frac{%
i}{%
\!\not\!%
q-\frac{1}{2}%
\!\not\!%
P-m}\Gamma ^{\mu \nu ...}\frac{i}{%
\!\not\!%
q+\frac{1}{2}%
\!\not\!%
P-m}\right\}  \nonumber \\
&&  \label{OrthoDec}
\end{eqnarray}
for parapositronium (pseudoscalar) and orthopositronium (vector, with
polarization $e_{\alpha }$) respectively . The $\Gamma ^{\mu \nu ...}$ is
the scattering amplitude for off-shell $e^{+}e^{-}$ with incoming momenta $%
\frac{1}{2}P+q$ and $\frac{1}{2}P-q$ into the final state $A$, $m$ is the
electron mass, $P$ the positronium four-momentum and $F_{B}=F_{B}\left(
q^{2},q\cdot P\right) $. From these amplitudes, the decay widths are
calculated as 
\[
\Gamma \left( ^{2J+1}S\rightarrow A\right) =\frac{1}{2J+1}\frac{1}{2M}\int
d\Phi _{A}\sum_{pol}\left| \mathcal{M}^{\mu \nu ...}\left(
^{2J+1}S\rightarrow A\right) \varepsilon _{\mu }^{*}\varepsilon _{\nu
}^{*}...\right| ^{2} 
\]
with $M$ the positronium mass.

What we have shown in our previous article \cite{OurWork1} is that when one
expresses the loop integration in (\ref{ParaDec}) or (\ref{OrthoDec}) via a
dispersion relation (\cite{Nishi}, \cite{kniehl}), one ends up with a
factorized convolution-type amplitude, i.e. for parapositronium: 
\begin{equation}
\mathcal{M}^{\mu \nu ...}\left( ^{1}S\rightarrow A\right) =\frac{C}{2}\int 
\frac{d^{3}\mathbf{k}}{\left( 2\pi \right) ^{3}2E_{\mathbf{k}}}\psi \left( 
\mathbf{k}^{2}\right) Tr\left\{ \gamma _{5}\left( -%
\!\not\!%
k^{\prime }+m\right) \Gamma ^{\mu \nu ...}\left( k,k^{\prime
},l_{1},...\right) \left( 
\!\not\!%
k+m\right) \right\}  \label{FinalSTD}
\end{equation}
with $\psi \left( \mathbf{k}^{2}\right) $ the bound state wavefunction, $%
\Gamma ^{\mu \nu ...}\left( k,k^{\prime },l_{1},...\right) $ the reduced
scattering amplitude for \textit{on-shell} $e^{+}e^{-}$ with momenta $k$ and 
$k^{\prime }$, $\mathbf{k}=-\mathbf{k}^{\prime }$ and $C=\sqrt{M}/m$ ($M$ is
the positronium mass). The analogous form for orthopositronium decay is
obtained using the substitution $\gamma ^{5}\rightarrow 
\!\not\!%
e$. The form factor $F_{B}$ is related with $\psi \left( \mathbf{k}%
^{2}\right) $ through 
\[
F_{B}=C\phi _{o}\mathcal{F}\left( \mathbf{k}^{2}\right) \left( \mathbf{k}%
^{2}+\gamma ^{2}\right) 
\]
with $\phi _{o}\mathcal{F}\left( \mathbf{k}^{2}\right) =\psi \left( \mathbf{k%
}^{2}\right) $, $\phi _{o}$ being the Schr\"{o}dinger wavefunction at the
origin, and $\gamma ^{2}=m^{2}-M^{2}/4$ (related to the binding energy and
fine-structure constant through $E_{B}=M-2m=-m\alpha ^{2}/4$).

\section{Paradimuonium Decay into $\gamma e^{+}e^{-}$}

The present section concerns paradimuonium, the singlet $\mu ^{+}\mu ^{-}$
bound state \cite{Dimuonium}. This state has not been observed yet. The
reason to consider the decay $p$-$Dm\rightarrow \gamma e^{+}e^{-}$ is that
it is the simplest process where a photon is kinematically allowed to have a
vanishing energy. Therefore, we will be able to test whether the proposed
factorization procedure (\ref{FinalSTD}) gives a correct analytical
behaviour to the amplitude in the soft photon limit. The simplicity of the
process comes from the pseudoscalar nature of the paradimuonium, which
allows a manifestly gauge invariant treatment throughout. Note that the
positronium decay $p$-$Ps\rightarrow \gamma \nu \overline{\nu }$ via a $%
Z^{0} $ has the same dynamics. After having analyzed this simple decay, we
will be ready to tackle the more interesting process $o$-$Ps\rightarrow
\gamma \gamma \gamma $.

The decay $p$-$Dm\rightarrow e^{+}e^{-}\gamma $ is shown in figure 1. When
the form factor $F_{B}$ appearing at the vertex $p$-$Dm\rightarrow \mu
^{+}\mu ^{-}$ allows a change of variable $q\rightarrow -q$, the two
amplitudes can be combined into 
\begin{equation}
\mathcal{M}\left( p\text{-}Dm\rightarrow e^{+}e^{-}\gamma \right)
=8me^{3}\varepsilon ^{\mu \nu \rho \sigma }k_{\rho }t_{\sigma }\varepsilon
_{\mu }^{*}\left( k\right) \frac{\left\{ \overline{u}\left( p\right) \gamma
_{\nu }v\left( p^{\prime }\right) \right\} }{t^{2}+i\varepsilon }\mathcal{I}%
\left( M^{2},x\right)  \label{TensorFactor}
\end{equation}
with $x=2\omega /M$ the reduced photon energy, $M$ the dimuonium mass and $m$
the muon mass.\ The loop integral, which can be viewed as an effective form
factor, is given by 
\[
\mathcal{I}\left( P^{2},x\right) =\eta \int \frac{d^{4}q}{\left( 2\pi
\right) ^{4}}F_{B}\frac{1}{\left( q-\frac{1}{2}P\right) ^{2}-m^{2}}\frac{1}{%
\left( q+\frac{1}{2}P\right) ^{2}-m^{2}}\frac{1}{\left( q-\frac{1}{2}%
P+k\right) ^{2}-m^{2}} 
\]
where $\eta =P^{2}/M^{2}$ and $t^{2}=P^{2}(1-x)$. Gauge invariance is
ensured by the factorized tensor structure, i.e. by the antisymmetric
Levi-Civita tensor. $\mathcal{I}\left( P^{2},x\right) $ is given by the same
expression as the two-photon decay integral we encountered previously \cite
{OurWork1}, the only difference being that $t^{2}\neq 0$.

The differential width is given by 
\[
\frac{d\Gamma \left( p\text{-}Dm\rightarrow e^{+}e^{-}\gamma \right) }{dx}=%
\frac{16\alpha ^{3}}{3}m^{2}\,M^{3}\left| \mathcal{I}\left( M^{2},x\right)
\right| ^{2}\rho \left( x,a\right) 
\]
with the phase space factor 
\[
\rho \left( x,a\right) =\sqrt{1-\frac{a}{1-x}}\left[ a+2\left( 1-x\right)
\right] \frac{x^{3}}{\left( 1-x\right) ^{2}} 
\]
where $a=4m_{e}^{2}/M^{2}$, and the bounds on $x$ are $\left[ 0,1-a\right] $.

We are going to calculate the integral $\mathcal{I}\left( M^{2},x\right) $
using dispersion relations. The main difference with the two-photon decay
case is the appearance of two different cuts (figure 2). We have shown \cite
{OurWork1} that considering the vertical cuts is strictly equivalent to the
decay amplitude calculation done using formula (\ref{FinalSTD}) where $%
\Gamma ^{\mu }$ is the scattering amplitude $\mu ^{+}\left( k^{\prime
}\right) \mu ^{-}\left( k\right) \rightarrow \gamma \gamma ^{*}\rightarrow
e^{+}e^{-}\gamma $ with on-shell muons. Standard approaches found in the
literature are then obtained by making an $\mathcal{O}\left( \alpha
^{2}\right) $ approximation in (\ref{FinalSTD}). Obviously, those approaches
completely miss the oblique cuts. What we are going to show is that those
forgotten cuts contribute at the order of $\gamma ^{2}=m^{2}-M^{2}/4\approx
\alpha ^{2}/4$, and more importantly, that \textit{the amplitude for }$p$%
\textit{-}$Dm\rightarrow e^{+}e^{-}\gamma $\textit{\ has a correct
soft-photon limit behaviour only if we take all the cuts into account}. By
this we mean that each cut gives a contribution to the amplitude that
behaves as a constant when the photon energy goes to zero. The combination
of the vertical and oblique cuts, on the contrary, forces the amplitude to
vanish in that limit. We thus recover the analytical behaviour expected from
Low's theorem \cite{Low} for the decay of $p$-$Dm\rightarrow \gamma \gamma
^{*}$ involving only neutral bosons (for $p$-$Dm\rightarrow \gamma
e^{+}e^{-} $, infrared divergences cannot arise from electron bremsstrahlung
processes due to selection rules).

\subsection{Soft photon limit property of the absorptive part}

To keep the discussion as general as possible, we extract the imaginary part
of $\mathcal{I}\left( P^{2},t^{2}\right) $ without specifying the form
factor $F_{B}$. The absorptive part is found by cutting the relevant
propagators as 
\begin{eqnarray*}
\func{Im}\mathcal{I}_{1}\left( P^{2},x\right) &=&\eta \int \frac{d^{4}q}{%
2\left( 2\pi \right) ^{4}}F_{B}\frac{2\pi i\delta \left( \left( q-\frac{1}{2}%
P\right) ^{2}-m^{2}\right) 2\pi i\delta \left( \left( q+\frac{1}{2}P\right)
^{2}-m^{2}\right) }{\left( q-\frac{1}{2}P+k\right) ^{2}-m^{2}} \\
\func{Im}\mathcal{I}_{2}\left( P^{2},x\right) &=&\eta \int \frac{d^{4}q}{%
2\left( 2\pi \right) ^{4}}F_{B}\frac{2\pi i\delta \left( \left( q+\frac{1}{2}%
P\right) ^{2}-m^{2}\right) 2\pi i\delta \left( \left( q-\frac{1}{2}%
P+k\right) ^{2}-m^{2}\right) }{\left( q-\frac{1}{2}P\right) ^{2}-m^{2}}
\end{eqnarray*}
for the vertical and oblique cuts, respectively. By a straightforward
integration, the first expression gives ($s=P^{2}$): 
\begin{equation}
\func{Im}\mathcal{I}_{1}\left( s,x\right) =\frac{\eta }{s}\frac{F_{B}\left(
q_{0}=0,\left| \mathbf{q}\right| =\sqrt{s/4-m^{2}}\right) }{16\pi x}\ln
\left[ \frac{1+\sqrt{1-4m^{2}/s}}{1-\sqrt{1-4m^{2}/s}}\right] \theta \left(
s-4m^{2}\right)  \label{Cut1Imaginary}
\end{equation}
while the second one cannot be completely integrated without specifying $%
F_{B}$: 
\begin{equation}
\func{Im}\mathcal{I}_{2}\left( s,x\right) =\frac{\eta }{s}\frac{1}{16\pi x}%
\int_{q_{\min }}^{q_{\max }}dq_{0}\frac{F_{B}\left( q_{0},\left| \mathbf{q}%
\right| =\sqrt{q_{0}^{2}+q_{0}\sqrt{s}+s/4-m^{2}}\right) }{q_{0}}\theta
\left( s-\frac{4m^{2}}{1-x}\right)  \label{Cut2Imaginary}
\end{equation}
with the bounds given by 
\begin{equation}
q_{\min }=-\frac{x\sqrt{s}}{4}\left( 1+\sqrt{1-\frac{4m^{2}}{s}\frac{1}{1-x}}%
\right) ,q_{\max }=-\frac{x\sqrt{s}}{4}\left( 1-\sqrt{1-\frac{4m^{2}}{s}%
\frac{1}{1-x}}\right)  \label{DispEEGBounds}
\end{equation}

Interestingly, the second cuts contribute for $q_{0}\neq 0$. This is in
sharp contrast with the two-photon decay, since there only the first cuts
exist. Further, approximated bound state wavefunction where a $\delta \left(
q_{0}\right) $ appears cannot be used (\cite{OrthoCorr}, \cite{Adkins}, \cite
{OrthoCorr2}), and one should revert to the full four-dimensional
Bethe-Salpeter wavefunction (for example the Barbieri-Remiddi one \cite
{BarbRem}).

Let us now demonstrate an important point, i.e. that in the soft photon
limit, the combination $\func{Im}\mathcal{I}_{1}\left( s,x\right) +\func{Im}%
\mathcal{I}_{2}\left( s,x\right) $ behaves as a constant when $x\rightarrow
0 $ despite the fact that each cut diverges in that limit. This will
guarantee that the imaginary part of the whole amplitude vanishes in the
soft photon limit thanks to the presence of $k_{\rho }$ in the tensor
structure of (\ref{TensorFactor}). This behaviour is what one expects from
Low's theorem \cite{Low}. For all the form factors we considered, the
behaviour of the combination $\func{Im}\mathcal{I}_{1}\left( s,x\right) +%
\func{Im}\mathcal{I}_{2}\left( s,x\right) $ when $x\rightarrow 0$ is
summarized by : 
\[
\func{Im}\mathcal{I}_{1}\left( s,x\right) +\func{Im}\mathcal{I}_{2}\left(
s,x\right) \stackrel{x\rightarrow 0}{\rightarrow }\frac{\eta }{s}\frac{1}{%
16\pi }\left[ \frac{F_{B}\left( 0\right) }{\sqrt{1-\frac{4m^{2}}{s}}}+\frac{%
\sqrt{s-4m^{2}}}{2}\frac{\partial F_{B}}{\partial q}\left( 0\right) \right] 
\]
Specifically, the form factors of interest are constructed from the
Schr\"{o}dinger wavefunction (see \cite{OurWork1}). Let us recall their
definition and give the limit obtained when using each of them : 
\begin{eqnarray}
F_{B}^{I} &\equiv &C\phi _{o}\frac{8\pi \gamma }{\mathbf{q}^{2}+\gamma ^{2}}
\label{FIBLimit} \\
&\Rightarrow &\func{Im}\mathcal{I}_{1}\left( s,x\right) +\func{Im}\mathcal{I}%
_{2}\left( s,x\right) \stackrel{x\rightarrow 0}{\rightarrow }\frac{\eta }{s}%
\frac{C\phi _{o}}{2}\gamma \frac{\left( s/4-\gamma ^{2}-m^{2}\right) }{\sqrt{%
1-\dfrac{4m^{2}}{s}}\left( \gamma ^{2}-m^{2}+\dfrac{s}{4}\right) ^{2}} 
\nonumber
\end{eqnarray}
\begin{eqnarray}
F_{B}^{II} &\equiv &C\phi _{o}\frac{32\pi \gamma ^{3}}{\left( \mathbf{q}%
^{2}+\gamma ^{2}\right) ^{2}}  \label{FIIBLimit} \\
&\Rightarrow &\func{Im}\mathcal{I}_{1}\left( s,x\right) +\func{Im}\mathcal{I}%
_{2}\left( s,x\right) \stackrel{x\rightarrow 0}{\rightarrow }\frac{\eta }{s}%
\frac{C\phi _{o}}{2}\gamma ^{3}\frac{\left( 4\gamma ^{2}-3\left(
s-4m^{2}\right) \right) }{\sqrt{1-\dfrac{4m^{2}}{s}}\left( \gamma ^{2}-m^{2}+%
\dfrac{s}{4}\right) ^{3}}  \nonumber
\end{eqnarray}
i.e. constant limits. Remark that for a constant form factor $F_{B}=F_{B}^{%
\text{Const}}$, the limit is simply 
\[
\func{Im}\mathcal{I}_{1}\left( s,x\right) +\func{Im}\mathcal{I}_{2}\left(
s,x\right) \stackrel{x\rightarrow 0}{\rightarrow }\frac{\eta }{s}\frac{%
F_{B}^{\text{Const}}}{16\pi }\frac{1}{\sqrt{1-4m^{2}/s}} 
\]
It is easy to verify that this result is indeed what can be calculated from
the imaginary parts : \newline
$\func{Im}\mathcal{I}_{1}\left( s,x\right) +\func{Im}\mathcal{I}_{2}\left(
s,x\right) =$%
\[
\begin{array}{r}
\dfrac{\eta }{s}\dfrac{F_{B}^{\text{Const}}}{16\pi x}\left[ \ln \left[ \frac{%
1+\sqrt{1-\frac{4m^{2}}{s}}}{1-\sqrt{1-\frac{4m^{2}}{s}}}\right] \theta
\left( s-4m^{2}\right) -\ln \left[ \frac{1+\sqrt{1-\frac{4m^{2}}{s\left(
1-x\right) }}}{1-\sqrt{1-\frac{4m^{2}}{s\left( 1-x\right) }}}\right] \theta
\left( s-\frac{4m^{2}}{1-x}\right) \right]
\end{array}
\]

In conclusion, the absorptive part of the amplitude is seen to vanish. Thus,
we expect that the dispersive part will also vanish in that limit. The
oblique cuts, omitted from standard analyzes, are seen to be essential to
maintain good analytical properties of the amplitude.

In the appendix, the restoration of analyticity by a cancellation among the
vertical and oblique cuts is presented in the context of the kaon decay $%
K_{S}^{0}\rightarrow e^{+}e^{-}\gamma $ via a charged pion loop. In this
analysis, the form factor describing the $K\rightarrow \pi \pi $ vertex is
taken as a constant. This decay is interesting since there is a close
similarity between this hadronic decay process and the present QED bound
state decay. The main conclusion is that even if charged particles are
present in intermediate states, the amplitude has to vanish in the soft
photon limit, as expected from the fact that the decays $K_{S}^{0}%
\rightarrow \gamma \gamma ^{*}$ or $p$-$Dm\rightarrow \gamma \gamma ^{*}$
involve only neutral bosons. Further, this $K_{S}^{0}$ decay provides a
physically sensible process where one can analyze the constant form factor
assumption, since for such a loosely bound system as the $p$-$Dm$ a constant
form factor cannot be realistic.

We will now evaluate the contribution of each cut for the case of the
Schr\"{o}dinger momentum wavefunction form factor $F_{B}^{I}$ in (\ref
{FIBLimit} ). In doing so, we will gain a better understanding of the
analyticity restoration at the level of the differential decay rate for $p$-$%
Dm\rightarrow e^{+}e^{-}\gamma $. The purpose of the next sections is only
illustrative, and the discussion can be straightforwardly transcribed for
the advocated improved form factor $F_{B}^{II}$ (\ref{FIIBLimit}) (see\cite
{OurWork1}).

\subsection{The vertical cuts reproduce standard approach results}

To precisely show what happens when the oblique cuts are forgotten, we
compute here the rate with the vertical cuts only. The present derivation is
therefore equivalent to standard analyses.

For our computation, we take the first cut imaginary part given by (\ref
{Cut1Imaginary}). The real part is found through an unsubstracted dispersion
relation with $\eta =s/M^{2}$ (see \cite{OurWork1}, \cite{Nishi}, \cite
{kniehl}) : 
\begin{eqnarray}
\mathcal{I}_{1}\left( M^{2},x\right) &=&\func{Re}\mathcal{I}_{1}\left(
M^{2},x\right) =\frac{C\phi _{o}}{\pi M^{2}}\frac{2\gamma }{x}%
\int_{4m^{2}}^{\infty }\frac{ds}{\left( s-M^{2}\right) ^{2}}\ln \left[ \frac{%
1+\sqrt{1-4m^{2}/s}}{1-\sqrt{1-4m^{2}/s}}\right]  \nonumber \\
&=&\frac{C\phi _{o}}{M^{3}}\frac{1}{x}\left[ \frac{2}{\pi }\arctan \frac{M}{%
2\gamma }\right]  \label{Cut1Real}
\end{eqnarray}
where the first equality holds since $M^{2}<4m^{2}$. Remark that for $%
x\rightarrow 1$, the second cuts vanish (see \ref{Cut2Imaginary}) and (\ref
{Cut1Real}) reproduce the result obtained in \cite{OurWork1} for the
parapositronium two-photon decay.

Inserting the result for $\mathcal{I}_{1}$ into the expression of the
differential rate with $C^{2}=M/m^{2}$ and $\left| \phi _{o}\right|
^{2}=\alpha ^{3}m^{3}/8\pi $, we get: 
\[
\frac{d\Gamma \left( p\text{-}Dm\rightarrow \gamma e^{+}e^{-}\right) }{dx}=%
\frac{\alpha ^{6}m}{6\pi }\,\left( \frac{4m^{2}}{M^{2}}\right) \left| \frac{2%
}{\pi }\arctan \frac{M}{2\gamma }\right| ^{2}\frac{\rho \left( x,a\right) }{%
x^{2}} 
\]
Note that, independently of the value for $\gamma $, the spectrum is always
linear in $x$ due to the $1/x$ in (\ref{Cut1Real}). This is an incorrect
soft photon behaviour since Low's theorem requires a cubic spectrum instead 
\cite{AnsatzWork} (when the amplitude behaves as $x$, the rate $\sim x^{2}$
and an additional $x$ comes from phase-space) . By expanding the
differential rate around $\gamma =0$, we get 
\[
\frac{d\Gamma \left( p\text{-}Dm\rightarrow \gamma e^{+}e^{-}\right) }{dx}=%
\frac{\alpha ^{6}m}{6\pi }\left( 1-\frac{\alpha }{\pi }+\frac{1}{4}\alpha
^{2}-\frac{25}{96\pi }\alpha ^{3}+\frac{3}{64}\alpha ^{4}+\mathcal{O}\left(
\alpha ^{5}\right) \right) \frac{\rho \left( x,a\right) }{x^{2}} 
\]
As lowest order, we recover the standard result: 
\begin{equation}
\frac{d\Gamma \left( p\text{-}Dm\rightarrow \gamma e^{+}e^{-}\right) }{dx}=%
\frac{\alpha ^{6}m_{\mu }}{6\pi }\frac{\rho \left( x,a\right) }{x^{2}}
\label{ReuslDiffCut1}
\end{equation}
Note that the order $\alpha $ corrections disappear when using the form
factor $F_{B}^{II}$ (see \cite{OurWork1}). For completeness, the total rate
is simply : 
\[
\Gamma \left( p\text{-}Dm\rightarrow \gamma e^{+}e^{-}\right) =\frac{\alpha
^{6}m}{6\pi }F\left( a\right) \left( \frac{4m^{2}}{M^{2}}\right) \left| 
\frac{2}{\pi }\arctan \frac{M}{2\gamma }\right| ^{2} 
\]
with 
\[
F\left( a\right) =\left[ \frac{4}{3}\sqrt{1-a}\left( a-4\right) +2\log
\left( \frac{1+\sqrt{1-a}}{1-\sqrt{1-a}}\right) \right] 
\]

\subsection{The oblique cut contribution to the rate}

We have just seen that the vertical cuts suffice to reproduce the lowest
order decay rate. It is interesting to investigate how the oblique cuts can
restore the analytical behaviour of the spectrum without affecting this
lowest order evaluation.

The imaginary part corresponding to the oblique cuts is, from (\ref
{Cut2Imaginary}), 
\begin{eqnarray*}
\func{Im}\mathcal{I}_{2}\left( s,x\right) &=&\frac{\eta }{s}\frac{C\phi _{o}%
}{16\pi x}\int_{q_{\min }}^{q_{\max }}\frac{dq_{0}}{q_{0}}\frac{8\pi \gamma 
}{q_{0}^{2}+q_{0}\sqrt{s}+s/4-m^{2}+\gamma ^{2}}\theta \left( s-\frac{4m^{2}%
}{1-x}\right) \\
&=&\frac{\eta }{s}\frac{C\phi _{o}}{16\pi x}8\pi \gamma \left[ \mathcal{H}%
\left( q_{\max }\right) -\mathcal{H}\left( q_{\min }\right) \right] \theta
\left( s-\frac{4m^{2}}{1-x}\right)
\end{eqnarray*}
with $q_{\min }$, $q_{\max }$ given in (\ref{DispEEGBounds}) and 
\[
\mathcal{H}\left( q\right) =-\frac{2\sqrt{s}}{M\left( M^{2}-s\right) }\ln
\left[ \frac{M+\sqrt{s}+2q}{M-\sqrt{s}-2q}\right] +\frac{2}{M^{2}-s}\ln
\left[ \frac{\left( \sqrt{s}+2q\right) ^{2}-M^{2}}{q^{2}}\right] 
\]
We should now use an unsubstracted dispersion relation to get $\mathcal{I}%
_{2}\left( M^{2},x\right) $. This integral is quite complicated and not very
interesting for the present purpose. Instead, we revert to numerical
evaluation of the dispersion integral for $\mathcal{I}_{2}\left(
M^{2},x\right) =\func{Re}\mathcal{I}_{2}\left( M^{2},x\right) $ as a
function of $x$, and compare it to (\ref{Cut1Real}). In the figure 3, we
plot the vertical cuts contribution (dashed line) 
\[
x\cdot \frac{M^{3}}{C\phi _{o}}\mathcal{I}_{1}\left( M^{2},x\right) =\frac{2%
}{\pi }\arctan \frac{M}{2\gamma } 
\]
and the complete result (solid line) 
\begin{eqnarray*}
x\cdot \frac{M^{3}}{C\phi _{o}}\mathcal{I}\left( M^{2},x\right) &=&x\cdot 
\frac{M^{3}}{C\phi _{o}}\left[ \mathcal{I}_{1}\left( M^{2},x\right) +%
\mathcal{I}_{2}\left( M^{2},x\right) \right] \\
&=&\frac{2}{\pi }\arctan \frac{M}{2\gamma }+\frac{M\gamma }{2\pi }\int_{%
\frac{4m^{2}}{1-x}}^{+\infty }\frac{ds}{s-M^{2}}\left[ \mathcal{H}\left(
q_{\max }\left( x\right) \right) -\mathcal{H}\left( q_{\min }\left( x\right)
\right) \right]
\end{eqnarray*}
for $m=1,M=1.999$ (i.e. $\gamma \approx 0.03$) and for $m=1,M=1.99994$ ($%
\gamma \approx 0.0077$). For $x\rightarrow 0$, one can verify that $x\cdot 
\mathcal{I}\left( M^{2},x\right) \rightarrow 0$. The figures show that near
zero the two cuts interfere destructively in order to maintain a correct
analytical behaviour for the whole amplitude. Away from $x=0$, the oblique
cut contributions are strongly suppressed relatively to the vertical one,
and this suppression increases as $\gamma $ decreases ($\mathcal{I}%
_{2}\rightarrow 0$ when $\gamma \rightarrow 0$). As can be seen on the
graph, it is typically for $x\lesssim \gamma $ that the oblique cut
contributes. Therefore, we can summarize by giving a simple representation
of the different contributions to the amplitude. From the figure 3, one can
see that the behaviours of the vertical cuts, the oblique cuts and their
combination are quite precisely modelled as 
\[
\mathcal{I}_{1}\sim \frac{1}{x},\mathcal{I}_{2}\sim -\frac{\gamma /M}{%
x\left( x+\gamma /M\right) }\Rightarrow \mathcal{I}=\mathcal{I}_{1}+\mathcal{%
I}_{2}\sim \left( \frac{1}{x+\gamma /M}\right) 
\]
As a consequence, the spectrum behaves as 
\begin{equation}
\frac{d\Gamma \left( p\text{-}Dm\rightarrow \gamma e^{+}e^{-}\right) }{dx}%
\sim \left| \mathcal{I}\right| ^{2}\rho \left( x,a\right) \sim x^{3}\left( 
\frac{1}{x+\gamma /M}\right) ^{2}\sim x\left( \frac{x}{x+\gamma /M}\right)
^{2}  \label{ResultSpectrum}
\end{equation}
i.e. a linear spectrum when $\gamma \rightarrow 0$ (\ref{ReuslDiffCut1}),
and a $x^{3}$ spectrum when $x$ is small. The effect of $\gamma \neq 0$ is
therefore to soften the photon spectrum and slightly reduce the total width.
This behaviour is exactly what we postulated in a previous work \cite
{AnsatzWork}.

In conclusion, the oblique cuts have a small contribution to the decay rate
comparatively to the vertical cuts. However, their presence is essential to
guarantee the analytical properties of the amplitude expected from Low's
theorem. Further, when tackling $\mathcal{O}\left( \alpha ^{2}\right) $
corrections, one must include contributions from the oblique cuts.

\section{Orthopositronium Decay to three Photons}

We now turn to the interesting decay $o$-$Ps\rightarrow \gamma \gamma \gamma 
$. We start as usual with the loop model amplitude. Dispersion techniques
express that amplitude into separate contributions, arising from different
cuts in figure 4. Since only off-shell intermediate leptons appear in the
loop, the complete amplitude has a correct soft photon behaviour. This
implies that one must consider all the cuts to preserve the analytical
properties of the amplitude. This is one conclusion of the previous section.

Let us give the loop model amplitude 
\begin{eqnarray}
\mathcal{M}^{\mu \nu \rho }\left( o\text{-}Ps\rightarrow \gamma \gamma
\gamma \right) &=&e^{3}e_{\alpha }\left( P\right) \int \frac{d^{4}q}{\left(
2\pi \right) ^{4}}F_{B}Tr\left\{ \gamma ^{\alpha }\frac{1}{%
\!\not\!%
q-\frac{1}{2}%
\!\not\!%
P-m}\Gamma ^{\mu \nu \rho }\frac{1}{%
\!\not\!%
q+\frac{1}{2}%
\!\not\!%
P-m}\right\}  \nonumber \\
&&  \label{LoopOrtho.}
\end{eqnarray}
where the vertex is given by six amplitudes grouped as 
\[
\Gamma ^{\mu \nu \rho }=\Gamma _{1}^{\mu \nu \rho }+\Gamma _{2}^{\mu \nu
\rho }+\Gamma _{3}^{\mu \nu \rho } 
\]
with 
\begin{eqnarray*}
\Gamma _{1}^{\mu \nu \rho } &=&\gamma ^{\nu }\frac{1}{%
\!\not\!%
q-\frac{1}{2}%
\!\not\!%
P+%
\!\not\!%
l_{2}-m}\gamma ^{\rho }\frac{1}{%
\!\not\!%
q+\frac{1}{2}%
\!\not\!%
P-%
\!\not\!%
l_{1}-m}\gamma ^{\mu } \\
&&+\gamma ^{\mu }\frac{1}{%
\!\not\!%
q-\frac{1}{2}%
\!\not\!%
P+%
\!\not\!%
l_{1}-m}\gamma ^{\rho }\frac{1}{%
\!\not\!%
q+\frac{1}{2}%
\!\not\!%
P-%
\!\not\!%
l_{2}-m}\gamma ^{\nu } \\
\Gamma _{2}^{\mu \nu \rho } &=&\gamma ^{\nu }\frac{1}{%
\!\not\!%
q-\frac{1}{2}%
\!\not\!%
P+%
\!\not\!%
l_{2}-m}\gamma ^{\mu }\frac{1}{%
\!\not\!%
q+\frac{1}{2}%
\!\not\!%
P-%
\!\not\!%
l_{3}-m}\gamma ^{\rho } \\
&&+\gamma ^{\rho }\frac{1}{%
\!\not\!%
q-\frac{1}{2}%
\!\not\!%
P+%
\!\not\!%
l_{3}-m}\gamma ^{\mu }\frac{1}{%
\!\not\!%
q+\frac{1}{2}%
\!\not\!%
P-%
\!\not\!%
l_{2}-m}\gamma ^{\nu } \\
\Gamma _{3}^{\mu \nu \rho } &=&\gamma ^{\mu }\frac{1}{%
\!\not\!%
q-\frac{1}{2}%
\!\not\!%
P+%
\!\not\!%
l_{1}-m}\gamma ^{\nu }\frac{1}{%
\!\not\!%
q+\frac{1}{2}%
\!\not\!%
P-%
\!\not\!%
l_{3}-m}\gamma ^{\rho } \\
&&+\gamma ^{\rho }\frac{1}{%
\!\not\!%
q-\frac{1}{2}%
\!\not\!%
P+%
\!\not\!%
l_{3}-m}\gamma ^{\nu }\frac{1}{%
\!\not\!%
q+\frac{1}{2}%
\!\not\!%
P-%
\!\not\!%
l_{1}-m}\gamma ^{\mu }
\end{eqnarray*}
and $P=l_{1}+l_{2}+l_{3}$. The amplitudes in $\Gamma _{1}^{\mu \nu \rho }$
are shown on figure 4. Using charge-conjugation, we can show that the two
drawn amplitudes are equal. The proof of this is in close analogy with the
demonstration of Furry's theorem, and requires a change of integration
variable $q\rightarrow -q$. This change is allowed for the form factor (see (%
\ref{FIBLimit}), (\ref{FIIBLimit})). Therefore, one can forget one term in
each $\Gamma _{i}^{\mu \nu \rho }$ and multiply the other by $2$. In the
following, to keep the combinational as clear as possible, we will continue
to consider all the six diagrams.

Let us now review the contributions aarising from the possible cuts in
figure 4.

\subsection{The vertical cut reproduces standard approach results}

First we have the six vertical cuts shown on figure 5. As demonstrated in
the paper \cite{OurWork1}, those vertical cuts reproduce the known result:
their combination is what is calculated with (\ref{FinalSTD}), i.e. at
lowest order \cite{OrePowell} 
\[
\Gamma \left( o\text{-}Ps\rightarrow \gamma \gamma \gamma \right) =\frac{%
2\left( \pi ^{2}-9\right) }{9\pi }\alpha ^{6}m 
\]
This result is obtained in the static limit, i.e. with $F_{B}\propto \delta
^{\left( 3\right) }\left( \mathbf{q}\right) $.

If this limit is not taken, one has to go through the integration of the
convolution-type amplitude (\ref{FinalSTD}). As explained in \cite{OurWork1}%
, the order of the corrections that will arise from that integration depends
on the explicit form for $F_{B}$. Let us recall that for the parapositronium
decay to two photons, we found that the form factors (\ref{FIBLimit}) led to 
$\mathcal{O}\left( \alpha \right) $ corrections and higher, while (\ref
{FIIBLimit}) gave corrections starting at $\mathcal{O}\left( \alpha
^{2}\right) $.

Finally, it is important to note that the six vertical cuts are separately
gauge invariant since the scattering amplitude for on-shell $%
e^{+}e^{-}\rightarrow \gamma \gamma \gamma $ is.

\subsection{The oblique cuts and the structure dependent contributions}

The oblique cuts are depicted in figure 6 for the photon $1$ on the
positronium side. Similar cuts for the photon $2$ or $3$ singled out are
easily drawn. The oblique cuts are not gauge invariant, contrary to the
vertical ones. To see this, it suffices to note that each of these oblique
cuts is, from the optical theorem, a product of a scattering amplitude $%
e^{+}e^{-}\rightarrow \gamma \gamma $ times a bremsstrahlung amplitude $o$-$%
Ps\rightarrow e^{+}e^{-}\gamma $. Gauge invariance is broken by the
bremsstrahlung amplitude, because the form factor is evaluated at different
momenta for different cuts. To visualize the situation, let us change the
momentum parametrization and draw amplitudes for $o$-$Ps\rightarrow
e^{+}e^{-}\gamma \left( l_{1}\right) $ as shown in figure 7. Obviously, when
contracted by $l_{1}^{\mu }$, the two amplitudes fail to cancel each other,
due to the different momentum dependences of $F_{B}$ : 
\begin{eqnarray*}
\mathcal{M}_{Brem1}^{\mu } &=&eF_{B}\left( p_{1}+l_{1},p_{2}\right) \left\{ 
\overline{u}\left( p_{1}\right) \gamma ^{\mu }\frac{i}{%
\!\not\!%
p_{1}+%
\!\not\!%
l_{1}-m}\gamma ^{\alpha }v\left( p_{2}\right) \right\} e_{\alpha }\left(
P\right) \\
&&+eF_{B}\left( p_{1},p_{2}+l_{1}\right) \left\{ \overline{u}\left(
p_{1}\right) \gamma ^{\alpha }\frac{i}{-%
\!\not\!%
p_{2}-%
\!\not\!%
l_{1}-m}\gamma ^{\mu }v\left( p_{2}\right) \right\} e_{\alpha }\left(
P\right) \\
&=&ieF_{B}\left( p_{1}+l_{1},p_{2}\right) \left\{ \overline{u}\left(
p_{1}\right) \frac{2p_{1}^{\mu }+\gamma ^{\mu }%
\!\not\!%
l_{1}}{2p_{1}\cdot l_{1}}\gamma ^{\alpha }v\left( p_{2}\right) \right\}
e_{\alpha }\left( P\right) \\
&&+ie^{2}F_{B}\left( p_{1},p_{2}+l_{1}\right) \left\{ \overline{u}\left(
p_{1}\right) \gamma ^{\alpha }\frac{-2p_{2}^{\mu }-%
\!\not\!%
l_{1}\gamma ^{\mu }}{2p_{2}\cdot l_{1}}v\left( p_{2}\right) \right\}
e_{\alpha }\left( P\right)
\end{eqnarray*}
and when contracted by $l_{1,\mu }$ : 
\[
l_{1,\mu }\mathcal{M}_{Brem1}^{\mu }=ie\left[ F_{B}\left(
p_{1}+l_{1},p_{2}\right) -F_{B}\left( p_{1},p_{2}+l_{1}\right) \right]
\left\{ \overline{u}\left( p_{1}\right) \gamma ^{\alpha }v\left(
p_{2}\right) \right\} e_{\alpha }\left( P\right) 
\]
Therefore, one should add to these diagrams the structure dependent
amplitude $\mathcal{M}_{SD1}^{\mu }$(figure 7). The new direct coupling
between $o$-$Ps,e^{+},e^{-}$ and $\gamma $ must be such that the combination
of the bremsstrahlung graphs and this structure graph is gauge invariant.
For the paradimuonium, we have not considered such structure terms because
the amplitude was gauge invariant throughout thanks to the factorized tensor
structure (see (\ref{TensorFactor})). This is a peculiar feature of the
two-photon decay of pseudoscalar positronium or dimuonium state.

Let us now turn to the soft photon behaviour. The structure terms do not
alter the cancellation among cuts in the soft photon limit (see the
corresponding discussion for $K_{S}^{0}$ decay in the appendix). Indeed,
Low's theorem applied to the bremsstrahlung plus structure dependent
amplitudes gives the following expansion around $l_{1}=0$ (see \cite{Low}, 
\cite{NonCstF}):

$\mathcal{M}_{Brem}^{\mu }+\mathcal{M}_{Struct}^{\mu }=$%
\[
iF_{B}\left( p_{1},p_{2}\right) \left\{ \overline{u}\left( p_{1}\right)
\left[ \frac{2p_{1}^{\mu }+\gamma ^{\mu }%
\!\not\!%
l_{1}}{2p_{1}\cdot l_{1}}\gamma ^{\alpha }-\gamma ^{\alpha }\frac{%
2p_{2}^{\mu }+%
\!\not\!%
l_{1}\gamma ^{\mu }}{2p_{2}\cdot l_{1}}\right] v\left( p_{2}\right) \right\}
e_{\alpha }\left( P\right) +\mathcal{O}\left( l_{1}\right) 
\]
i.e. terms of order $\left( l_{1}\right) ^{0}$ cancel because $\mathcal{M}%
_{Brem}^{\mu }+\mathcal{M}_{Struct}^{\mu }$ is gauge invariant. Importantly,
note that the form factor is now evaluated at $\left(
p_{1}^{2}=m^{2},p_{2}^{2}=m^{2}\right) $, i.e. at the same point as for the
vertical cuts. Since for a constant form factor the combination of all the
cuts necessarily behaves correctly in the soft photon limit, and since a
momentum dependence in the form factor only modifies $\mathcal{O}\left(
l_{1}\right) $ terms, the conclusion follows.

Remark that the appearance of these new structure contributions could have
been guessed from the start, since the loop model amplitude (\ref{LoopOrtho.}%
) fails to be gauge invariant due to momentum dependences in the form
factor. More precisely, the Ward identity $l_{1,\mu }\mathcal{M}^{\mu \nu
\rho }\left( o\text{-}Ps\rightarrow \gamma \gamma \gamma \right) =0$ is
verified only if the form factor allows for linear changes of the
integration variable like $q\rightarrow q+l_{1}$. A general form factor will
not allow such shifts and one must supplement the loop model with new
structure dependent amplitudes, as noted in \cite{AnsatzWork}. The
dispersion technique we followed here is interesting since we get a more
precise information on their origin, and we are able to constraint them
through Low's theorem.

The same discussion can be made for the photons $2$ and $3$. One ends up
with twelve oblique cuts, to which six structure dependent amplitudes must
be added. All these new contributions should become important at the order $%
\gamma ^{2}\approx \alpha ^{2}/4$, so that one can really question the
completeness of the results given in the literature \cite{Bterm}.

\section{Conclusions}

We have shown that the standard approaches used to calculate positronium
decay rates cannot be used to fully evaluate $\mathcal{O}\left( \alpha
^{2}\right) $ corrections. In the case of orthopositronium, many new
contributions arise at that order from processes where the electron emerging
from the bound state decay is off-shell. Those contributions are essential
to preserve the basic property of analyticity of the positronium decay
amplitude, i.e. its vanishing in the soft photon limit. Further, gauge
invariance requires some interplay between the annihilation process and the
bound state dynamics through structure dependent amplitudes.

All these new contributions vanish in the static limit, i.e. when the form
factor is replaced by a delta function $\delta ^{\left( 3\right) }\left( 
\mathbf{q}\right) $. Therefore, one can view these as non-perturbative
effects arising from the binding of the constituents in the bound state.
Indeed, the size of the corrections is fixed by the $\gamma ^{2}$ appearing
in the bound state wavefunction. Concerning this wavefunction, forms where
the dependence on the energy $q_{0}$ is replaced by a $\delta \left(
q_{0}\right) $ should not be used because it is specifically for non-zero
value of the energy that the additional diagrams contribute. This raises the
question of the form for $F_{B}$ to be used. Let us recall that in \cite
{OurWork1}, we advoquate the use of the improved form factor $F_{B}^{II}$
instead of the usual Schr\"{o}dinger momentum wavefunction.

In conclusion, it should now be clear that $\mathcal{O}\left( \alpha
^{2}\right) $ corrections as given in the literature are rather incomplete.
A lot of work is still needed before a definite theoretical prediction up to
that order can be compared to experiment. Even if the theoretical basis
seems settled, the orthopositronium life-time puzzle remains an open
question.\newline

{\Large Acknowledgements :} G. L. C. was partially supported by Conacyt
(M\'{e}xico) under contract No 32429. C. S. and S. T. acknowledge financial
supports from FNRS (Belgium).

\qquad

\qquad

\section{Appendix : Kaon Radiative Decay $K_{S}^{0}\rightarrow \gamma
e^{+}e^{-}$}

The amplitude for $K_{S}^{0}\rightarrow \gamma e^{+}e^{-}$ is computed at
lowest order in a pion loop model. The pion is treated as a point-like
charged particle, which allows simple scalar QED treatment. Special
attention is paid to the low energy behaviour of the amplitude.

\subsection{Decay Amplitude and Loop Integration}

The amplitude for kaon decay into $\gamma e^{+}e^{-}$ is given by 
\begin{eqnarray*}
\mathcal{M}\left( K_{S}^{0}\rightarrow \gamma e^{+}e^{-}\right) &=&-2ie^{3}%
\mathcal{M}\left( K_{S}^{0}\rightarrow \pi ^{+}\pi ^{-}\right) \varepsilon
_{\mu }^{*}\left( k\right) \frac{\left\{ \overline{u}\left( p\right) \gamma
_{\nu }v\left( p^{\prime }\right) \right\} }{t^{2}} \\
&&\times \int \frac{d^{d}q}{\left( 2\pi \right) ^{d}}\mathcal{M}^{\mu \nu
}\left( \pi ^{+}\pi ^{-}\rightarrow \gamma \gamma \right)
\end{eqnarray*}
with $t=p+p^{\prime }=P-k$. The $\pi ^{+}\pi ^{-}\rightarrow \gamma \gamma $
amplitude arises from the one-photon (figure 8a) and two-photon (the
so-called seagull graph, figure 8b) coupling amplitudes as 
\[
\mathcal{M}^{\mu \nu }\left( \pi ^{+}\pi ^{-}\rightarrow \gamma \gamma
\right) =\frac{\left( 2q-k\right) ^{\mu }\left( 2q+t\right) ^{\nu }-g^{\mu
\nu }\left( q^{2}-m^{2}\right) }{\left( \left( q+t\right) ^{2}-m^{2}\right)
\left( \left( q-k\right) ^{2}-m^{2}\right) \left( q^{2}-m^{2}\right) } 
\]
with $m$ the pion mass. The amplitude $\mathcal{M}\left(
K_{S}^{0}\rightarrow \pi ^{+}\pi ^{-}\right) $, taken as a constant, has
been factored out. The integration is done using dimensional regularization
to preserve gauge invariance and get a finite result. We obtain

$\dint \dfrac{d^{d}q}{\left( 2\pi \right) ^{d}}\mathcal{M}^{\mu \nu }\left(
\pi ^{+}\pi ^{-}\rightarrow \gamma \gamma \right) $%
\[
=\frac{-i}{\left( 4\pi \right) ^{2}}\int_{0}^{1}dx\int_{0}^{1-x}dy\left[ 
\tfrac{4xy}{\Delta }\left[ g^{\mu \nu }\left( k\cdot t\right) -t^{\mu
}k^{\nu }\right] +\tfrac{y\left( 1-2y\right) }{\Delta }g^{\mu \nu
}t^{2}\right] 
\]
The denominator function is $\Delta =m^{2}\left( 1-4\left( a-b\right)
xy+4by\left( y-1\right) \right) $ with the definitions $%
a=M^{2}/4m^{2},b=t^{2}/4m^{2}$ and $M$ the kaon mass. In the kaon
rest-frame, $b=M^{2}\left( 1-\omega \right) /4m^{2}$ with $\omega $ the
reduced photon energy $2k^{0}/M$. When integrating over Feynman parameters,
the last term vanishes. Therefore, the gauge invariance of the amplitude
becomes manifest:

$\mathcal{M}\left( K_{S}^{0}\rightarrow \gamma e^{+}e^{-}\right) =\dfrac{%
-2e^{3}}{\left( 4\pi \right) ^{2}}\mathcal{M}\left( K_{S}^{0}\pi ^{+}\pi
^{-}\right) \dfrac{1}{m^{2}}F\left( a,b\right) \times $%
\begin{equation}
\qquad \qquad \qquad \qquad \,\varepsilon _{\mu }^{*}\left( k\right) \left\{ 
\overline{u}\left( p\right) \gamma _{\nu }v\left( p^{\prime }\right)
\right\} \frac{g^{\mu \nu }\left( k\cdot t\right) -t^{\mu }k^{\nu }}{t^{2}}
\label{Amplitude}
\end{equation}
$F\left( a,b\right) $ is the Feynman parameter integral 
\[
F\left( a,b\right) =\int_{0}^{1}dy\int_{0}^{1-y}dx\frac{4xy}{1-4\left(
a-b\right) xy+4by\left( y-1\right) +i\varepsilon } 
\]
The prescription $i\varepsilon $ gives the sign of the imaginary part.

\subsection{Feynman Parameter Integration via Dispersion Relations}

To calculate $F\left( a,b\right) $, we shall use dispersion relations (see
for example \cite{Nishi}, \cite{kniehl}). A direct integration is possible,
but we shall gain insight into the dynamics of the process by using
dispersion techniques.

\paragraph{Absorptive Part Extraction\newline
\newline
}

A first integration over the Feynman parameters gives 
\begin{equation}
F\left( a,b\right) =\int_{0}^{1}dy\left[ \frac{y-1}{a-b}+\frac{\ln K\left(
b\right) -\ln K\left( a\right) }{4y\left( a-b\right) ^{2}}+\frac{b\left(
y-1\right) \left( \ln K\left( b\right) -\ln K\left( a\right) \right) }{%
\left( a-b\right) ^{2}}\right]  \label{FabElements}
\end{equation}
where $\ln K\left( x\right) =\ln \left( 1+4x\left( y-1\right) y-i\varepsilon
\right) $. The imaginary part of $F\left( a,b\right) $ comes from the values
of $y$ for which the argument of a logarithm is negative. There the
logarithm imaginary part is $-i\pi $. Integrating on the relevant $y$
values, we find 
\begin{eqnarray}
\func{Im}F\left( a,b\right) &=&\frac{\pi \left( \ln \left[ \sqrt{a}+\sqrt{a-1%
}\right] -b\sqrt{\frac{a-1}{a}}\right) }{2\left( a-b\right) ^{2}}\theta
\left( a-1\right)  \nonumber \\
&&-\frac{\pi \left( \ln \left[ \sqrt{b}+\sqrt{b-1}\right] -b\sqrt{\frac{b-1}{%
b}}\right) }{2\left( a-b\right) ^{2}}\theta \left( b-1\right)  \label{ImFab}
\end{eqnarray}
or, recalling the definition of $a$ and $b$ (see \cite{Sehgal}) 
\[
\func{Im}F\left( a,b\right) =\left\{ \frac{4\pi m^{4}}{M^{4}\omega ^{2}}\ln
\left[ \frac{1+\sqrt{1-\frac{4m^{2}}{M^{2}}}}{1-\sqrt{1-\frac{4m^{2}}{M^{2}}}%
}\right] -\frac{2\pi m^{2}}{M^{2}}\frac{1-\omega }{\omega ^{2}}\sqrt{1-%
\tfrac{4m^{2}}{M^{2}}}\right\} \theta \left( M^{2}-4m^{2}\right) 
\]
\begin{equation}
-\left\{ \frac{4\pi m^{4}}{M^{4}\omega ^{2}}\ln \left[ \frac{1+\sqrt{1-\frac{%
4m^{2}}{M^{2}\left( 1-\omega \right) }}}{1-\sqrt{1-\frac{4m^{2}}{M^{2}\left(
1-\omega \right) }}}\right] -\frac{2\pi m^{2}}{M^{2}}\frac{1-\omega }{\omega
^{2}}\sqrt{1-\tfrac{4m^{2}}{M^{2}\left( 1-\omega \right) }}\right\} \theta
\left( M^{2}-\frac{4m^{2}}{1-\omega }\right)  \label{ImFws}
\end{equation}

Let us consider the figure $9$. Obviously, the first line in (\ref{ImFws})
corresponds to the vertical cuts (which contribute only if the decaying
particle's mass $M^{2}$ is greater than $\left( 2m\right) ^{2}$) while the
second line corresponds to the oblique cuts (which contribute only if the
virtual photon energy $t^{2}$ is greater than $\left( 2m\right) ^{2}$).

Further, one can see that in the soft photon limit $\omega \rightarrow 0$
(or $b\rightarrow a$), $\func{Im}F\left( a,b\right) $ behaves as a constant.
Indeed, applying L'Hospital's rule twice, we get 
\begin{equation}
\func{Im}F\left( \frac{M^{2}}{4m^{2}},\frac{M^{2}\left( 1-\omega \right) }{%
4m^{2}}\right) \stackrel{\omega \rightarrow 0}{=}\pi \frac{2m^{4}}{M^{4}}%
\frac{1}{\sqrt{1-4m^{2}/M^{2}}}  \label{ImSoftLim}
\end{equation}
while individually the contribution of each cut is divergent as $%
1/(a-b)^{2}\sim 1/\omega ^{2}$. We will come back to this point later.

\paragraph{Dispersive Part Integral\newline
\newline
}

In order to write down the unsubstracted dispersion integral, let us express 
$\func{Im}F\left( a,b\right) $ as a function of the available energy through 
$a\left( s\right) =s/4m^{2}$ and $b\left( s\right) =s\left( 1-\omega \right)
/4m^{2}$. Then 
\begin{equation}
\func{Re}F\left( a\left( s_{o}\right) ,b\left( s_{o}\right) \right) =\frac{P%
}{\pi }\int \frac{ds}{s-s_{o}}\func{Im}F\left( a\left( s\right) ,b\left(
s\right) \right)  \label{DispReFab}
\end{equation}
with $s_{o}<4m^{2}$, such that we can omit the principal part. For such a
kinematics $F\left( a\left( s_{o}\right) ,b\left( s_{o}\right) \right) =%
\func{Re}F\left( a\left( s_{o}\right) ,b\left( s_{o}\right) \right) $. In
the next section we shall analytically continue $F$ to the physical value $%
s_{o}=M^{2}$. The result of (\ref{DispReFab}) is easily obtained in terms of
the integrals 
\[
\int_{0}^{1}\frac{dy}{y_{o}-y}\ln \left[ \tfrac{1+\sqrt{1-y}}{1-\sqrt{1-y}}%
\right] =2\arcsin ^{2}\sqrt{\frac{1}{y_{o}}}\text{, }\int_{0}^{1}\frac{dy}{%
y_{o}-y}\sqrt{1-y}=2-2\sqrt{y_{o}-1}\arcsin \sqrt{\frac{1}{y_{o}}} 
\]
valid for $y_{o}>1$ (see \cite{Nishi}). We can write $F\left( a,b\right) $
with $0<a<1$ and $0<b<1$ as 
\begin{equation}
F\left( a,b\right) =-\frac{1}{2\left( a-b\right) }+\frac{1}{\left(
a-b\right) ^{2}}\left( \frac{1}{2}\left( f\left( a\right) -f\left( b\right)
\right) +b\left( g\left( a\right) -g\left( b\right) \right) \right)
\label{ResultfgFab}
\end{equation}
in terms of $f\left( x\right) \equiv \arcsin ^{2}\sqrt{x}$ and $g\left(
x\right) \equiv \sqrt{\frac{1-x}{x}}\arcsin \sqrt{x}$.

\paragraph{Analytic Continuation\newline
\newline
}

The results for $a>1$ and $b>1$, are obtained by analytic continuation of (%
\ref{ResultfgFab}). The analytic continuation of $f\left( a\right) $ and $%
g\left( a\right) $ are 
\begin{eqnarray*}
f\left( x\right) &=&\arcsin ^{2}\left( \sqrt{x}\right) =-\left( \ln \left( 
\sqrt{x}+\sqrt{x-1}\right) -\frac{1}{2}i\pi \right) ^{2} \\
g\left( x\right) &=&\sqrt{\frac{1-x}{x}}\arcsin \sqrt{x}=\sqrt{\frac{x-1}{x}}%
\left( \ln \left( \sqrt{x}+\sqrt{x-1}\right) -\frac{1}{2}i\pi \right)
\end{eqnarray*}
The equalities hold for $x\in C,\func{Im}x>0$. The right hand sides define
the analytical continuation for $x>1$. Hence the result is conveniently
expressed as (\ref{ResultfgFab}) with 
\begin{eqnarray}
f\left( x\right) &=&\left\{ 
\begin{array}{ll}
\arcsin ^{2}\left( \sqrt{x}\right) & 0<x<1 \\ 
-\left( \ln \left( \sqrt{x}+\sqrt{x-1}\right) -\frac{1}{2}i\pi \right) ^{2}
& x>1
\end{array}
\right.  \label{ResultFab} \\
g\left( x\right) &=&\left\{ 
\begin{array}{ll}
\sqrt{\frac{1-x}{x}}\arcsin \left( \sqrt{x}\right) & 0<x<1 \\ 
\sqrt{\frac{x-1}{x}}\left( \ln \left( \sqrt{x}+\sqrt{x-1}\right) -\frac{1}{2}%
i\pi \right) & x>1
\end{array}
\right.  \nonumber
\end{eqnarray}
Had we chosen to analytically continue $f$ and $g$ in the lower half complex
plane, their imaginary parts would have had the opposite sign. One can
verify from (\ref{ResultFab}) that the upper half complex plane analytic
continuation reproduces $\func{Im}F\left( a,b\right) $ as given in (\ref
{ImFab}). This result corresponds to \cite{Bergstr} where a sign mistake has
to be corrected, and to \cite{twoPhot}, \cite{Pest2}.

\subsection{Decay Width, Differential Width and Low's Theorem}

The differential decay width is given by $d\Gamma \left(
K_{S}^{0}\rightarrow e^{+}e^{-}\gamma \right) =\frac{1}{2M}\sum_{spin}\left| 
\mathcal{M}\right| ^{2}d\Phi _{3}$. After a straightforward integration over
the electron-positron phase-space, we get the differential rate in terms of
the reduced photon energy $\omega $: 
\[
\frac{d\Gamma \left( K_{S}^{0}\rightarrow e^{+}e^{-}\gamma \right) /d\omega 
}{\Gamma \left( K_{S}^{0}\rightarrow \pi ^{+}\pi ^{-}\right) }=\frac{\alpha
^{3}}{3\pi ^{3}}\frac{\left| F\left( \frac{1}{a_{\pi }},\frac{1-\omega }{%
a_{\pi }}\right) \right| ^{2}}{a_{\pi }^{2}\sqrt{1-a_{\pi }}}\sqrt{1-\frac{%
a_{e}}{1-\omega }}\left[ a_{e}+2\left( 1-\omega \right) \right] \frac{\omega
^{3}}{\left( 1-\omega \right) ^{2}} 
\]
with $a_{e}=4m_{e}^{2}/M^{2}$ and $a_{\pi }=4m^{2}/M^{2}=1/a$. The bounds on
the $\omega $ integration are $\omega _{\min }=0$ and $\omega _{\max
}=1-a_{e}$. A numerical integration of the differential rate gives the
prediction for the rate $K_{S}^{0}\rightarrow e^{+}e^{-}\gamma $ relatively
to $K_{S}^{0}\rightarrow \pi ^{+}\pi ^{-}$%
\begin{equation}
R_{\pi ^{+}\pi ^{-}}=\frac{\Gamma \left( K_{S}^{0}\rightarrow
e^{+}e^{-}\gamma \right) }{\Gamma \left( K_{S}^{0}\rightarrow \pi ^{+}\pi
^{-}\right) }\simeq 4.70\times 10^{-8}  \label{Rpipi}
\end{equation}
where the real and imaginary part of $F\left( \frac{1}{a_{\pi }},\frac{%
1-\omega }{a_{\pi }}\right) $ contribute for $1.26\times 10^{-8}$ and $%
3.43\times 10^{-8}$, respectively.

Let us turn to the soft photon behaviour of the differential width. We have
to analyze the loop integral function $F\left( \frac{1}{a_{\pi }},\frac{%
1-\omega }{a_{\pi }}\right) $. This function tends to a constant for very
low $\omega $: 
\begin{equation}
F\left( \frac{1}{a_{\pi }},\frac{1-\omega }{a_{\pi }}\right) \stackrel{%
\omega \rightarrow 0}{\rightarrow }-\frac{a_{\pi }}{4}+\frac{a_{\pi }^{2}}{%
4\left( a_{\pi }-1\right) }g\left( \frac{1}{a_{\pi }}\right)
\label{SoftLimit}
\end{equation}
(note that for $a_{\pi }<1$, the imaginary part of (\ref{SoftLimit}) is the
same as in (\ref{ImSoftLim})). Let us emphasize the strong cancellation
between the two cuts in the soft photon limit. Individually, each cut
diverges as $\omega ^{2}$, but their combination is convergent. Since the
rest of the amplitude tends to zero like $\omega $ (due to the $\left[
g^{\mu \nu }\left( k\cdot t\right) -t^{\mu }k^{\nu }\right] $ factor), 
\textit{Low's theorem is verified: the amplitude behaves as }$\omega $%
\textit{\ near }$\omega =0$\textit{.} The resulting spectrum is thus in $%
\omega ^{3}$ ($\omega ^{2}$ from the squared amplitude and a $\omega $ from
phase space). In other words, if we had forgotten one cut, the resulting
spectrum behaviour would have been divergent as $1/\omega $ near zero
instead of vanishing like $\omega ^{3}$. In fact, this $\omega ^{3}$
spectrum is exactly the same as in the $\pi ^{0}\rightarrow e^{+}e^{-}\gamma 
$ differential decay rate.

\paragraph{Comparison with the Two-Photon Decay Mode\newline
\newline
}

From the amplitude (\ref{Amplitude}), we readily obtain the amplitude for $%
K_{S}^{0}\rightarrow \gamma \gamma $ by removing the electron current, the
photon propagator and by taking the limit $\omega \rightarrow 1$: 
\[
\mathcal{M}\left( K_{S}^{0}\rightarrow \gamma \gamma \right) =\frac{-2\alpha 
}{4\pi }\frac{\mathcal{M}\left( K_{S}^{0}\rightarrow \pi ^{+}\pi ^{-}\right) 
}{m^{2}}\varepsilon _{\mu }^{*}\left( k\right) \varepsilon _{\nu }^{*}\left(
k^{\prime }\right) \left[ g^{\mu \nu }\left( k\cdot k^{\prime }\right)
-k^{\prime \mu }k^{\nu }\right] F\left( \frac{M^{2}}{4m^{2}},0\right) 
\]
with $F\left( \frac{1}{a_{\pi }},0\right) =\frac{1}{2}a_{\pi }^{2}\left\{
f\left( \frac{1}{a_{\pi }}\right) -\frac{1}{a_{\pi }}\right\} $ i.e. 
\begin{eqnarray}
F\left( \tfrac{M^{2}}{4m^{2}},0\right) &=&\frac{4m^{4}}{M^{2}}\left( \frac{%
i\pi }{M^{2}}\ln \left[ \tfrac{1+\sqrt{1-\frac{4m^{2}}{M^{2}}}}{1-\sqrt{1-%
\frac{4m^{2}}{M^{2}}}}\right] -\frac{1}{2}\left( \frac{1}{m^{2}}-\frac{\pi
^{2}}{M^{2}}+\frac{1}{M^{2}}\ln ^{2}\left[ \tfrac{1+\sqrt{1-\frac{4m^{2}}{%
M^{2}}}}{1-\sqrt{1-\frac{4m^{2}}{M^{2}}}}\right] \right) \right)  \nonumber
\\
&&  \label{FTwoPhotons}
\end{eqnarray}
which corresponds to the result given in \cite{Nishi}, \cite{twoPhot}, \cite
{chch}. The relative decay rate is : 
\[
\frac{\Gamma \left( K_{S}^{0}\rightarrow \gamma \gamma \right) }{\Gamma
\left( K_{S}^{0}\rightarrow \pi ^{+}\pi ^{-}\right) }=\frac{\alpha ^{2}}{\pi
^{2}}\frac{1}{a_{\pi }^{2}\sqrt{1-a_{\pi }}}\left| F\left( \frac{1}{a_{\pi }}%
,0\right) \right| ^{2}\simeq 2.94\times 10^{-6} 
\]
while the experimental value for this ratio is $\left( 3.5\pm 1.3\right)
\times 10^{-6}$ \cite{PDG}.

We can now compare the two electromagnetic modes : 
\[
R_{\gamma \gamma }=\frac{\Gamma \left( K_{S}^{0}\rightarrow e^{+}e^{-}\gamma
\right) }{\Gamma \left( K_{S}^{0}\rightarrow \gamma \gamma \right) }\simeq
0.016 
\]
in accordance with \cite{twoPhot} and \cite{Sehgal}. This ratio $R_{\gamma
\gamma }$ is similar to \cite{PDG} 
\[
\frac{\Gamma \left( \pi ^{0}\rightarrow e^{+}e^{-}\gamma \right) }{\Gamma
\left( \pi ^{0}\rightarrow \gamma \gamma \right) }\simeq 0.012 
\]
The small difference is due to the phase-space factor. The similarity
between a constant coupling model like for $\pi ^{0}\rightarrow
e^{+}e^{-}\gamma $ and the present loop model can be understood from the
behaviour of the photon energy spectrum. Indeed, the pure phase-space
spectrum is very strongly peaked at high $\omega $, and therefore the decay
rate is quite insensible to the detail of the $F\left( \frac{1}{a_{\pi }},%
\frac{1-\omega }{a_{\pi }}\right) $ function. Specifically, if we replace $%
F\left( \frac{1}{a_{\pi }},\frac{1-\omega }{a_{\pi }}\right) $ by its value
for $\omega =1$ (\ref{FTwoPhotons}) we find the ratio $R_{\pi ^{+}\pi
^{-}}\simeq 4.671\times 10^{-8}$, very close to (\ref{Rpipi}).

\subsection{Soft Photon Behaviour with a non-constant Form Factor}

We now turn to the case of a non-constant amplitude $\mathcal{M}\left(
K_{S}^{0}\rightarrow \pi ^{+}\pi ^{-}\right) \equiv \mathcal{F}\left( \left(
q+k\right) ^{2},\left( q-t\right) ^{2}\right) $. We analyze the process from
the point of view of Low's theorem. The constraint of gauge invariance will
be seen to require some new structure dependent contributions. We will prove
that those new contributions do not affect the interference between
absorptive parts observed in previous sections, responsible for the
vanishing of the amplitude in the soft photon limit.

Let us concentrate on the absorptive part of the amplitude for $%
K_{S}^{0}\rightarrow \gamma e^{+}e^{-}$. Using the optical theorem (see
figure 2), we write 
\begin{equation}
2\func{Im}\mathcal{M}^{\mu }\left( K_{S}^{0}\rightarrow \gamma
e^{+}e^{-}\right) =\mathcal{V}^{\mu }+\mathcal{D}^{\mu }  \label{ImVD}
\end{equation}
with $\mathcal{V}^{\mu }$ the vertical cut contribution and $\mathcal{D}%
^{\mu }$ the diagonal one : 
\begin{eqnarray*}
\mathcal{V}^{\mu } &=&\int d\Phi _{\pi \pi }\,\left( 2\pi \right) ^{4}\delta
^{\left( 4\right) }\left( P-p_{1}-p_{2}\right) \,\mathcal{M}\left(
K_{S}^{0}\rightarrow \pi ^{+}\left( p_{1}\right) \pi ^{-}\left( p_{2}\right)
\right) \\
&&\qquad \times \mathcal{M}^{\mu }\left( \pi ^{+}\left( p_{1}\right) \pi
^{-}\left( p_{2}\right) \rightarrow \gamma \gamma ^{*}\rightarrow \gamma
e^{+}e^{-}\right) \\
\mathcal{D}^{\mu } &=&\int d\Phi _{\pi \pi }\,\,\left( 2\pi \right)
^{4}\delta ^{\left( 4\right) }\left( P-p_{1}-p_{2}-k\right) \,\mathcal{M}%
^{\mu }\left( K_{S}^{0}\rightarrow \pi ^{+}\left( p_{1}\right) \pi
^{-}\left( p_{2}\right) \gamma \left( k\right) \right) \\
&&\qquad \times \mathcal{M}\left( \pi ^{+}\left( p_{1}\right) \pi ^{-}\left(
p_{2}\right) \rightarrow \gamma ^{*}\rightarrow e^{+}e^{-}\right)
\end{eqnarray*}

The $\pi \pi \rightarrow \gamma e^{+}e^{-}$ amplitude is gauge invariant, as
can be easily verified, and so is $\mathcal{V}^{\mu }$, independently of the
form of $\mathcal{F}\left( p_{1}^{2},p_{2}^{2}\right) $. Therefore $\mathcal{%
D}^{\mu }$ has to be gauge invariant too. The two bremsstrahlung
contributions to $\mathcal{D}^{\mu }$ are 

$\mathcal{M}_{IB}^{\mu }\left( K_{S}^{0}\rightarrow \pi ^{+}\left(
p_{1}\right) \pi ^{-}\left( p_{2}\right) \gamma \left( k\right) \right) =$%
\[
\mathcal{F}\left( \left( p_{1}+k\right) ^{2},p_{2}^{2}\right) \frac{%
p_{1}^{\mu }}{p_{1}\cdot k}-\mathcal{F}\left( p_{1}^{2},\left(
p_{2}+k\right) ^{2}\right) \frac{p_{2}^{\mu }}{p_{2}\cdot k}
\]
Trivially, gauge invariance is verified if the form factor is constant.\
Problems arise if $\mathcal{F}\left( p_{1}+k,p_{2}\right) \neq \mathcal{F}%
\left( p_{1},p_{2}+k\right) $, since when contracted by $k^{\mu }$ the two
terms fail to cancel. To maintain gauge invariance, one must supplement the
amplitude $K_{S}^{0}\rightarrow \pi ^{+}\pi ^{-}\gamma $ with a structure
dependent term, for example $\mathcal{M}_{SD}^{\mu }\left(
K_{S}^{0}\rightarrow \pi ^{+}\left( p_{1}\right) \pi ^{-}\left( p_{2}\right)
\gamma \left( k\right) \right) =$%
\begin{equation}
\left[ \mathcal{F}\left( p_{1}^{2},\left( p_{2}+k\right) ^{2}\right) -%
\mathcal{F}\left( \left( p_{1}+k\right) ^{2},p_{2}^{2}\right) \right] \frac{%
\left( p_{1}-p_{2}\right) ^{\mu }}{p_{1}\cdot k-p_{2}\cdot k}  \label{SDTerm}
\end{equation}
such that $k^{\mu }\mathcal{M}_{IB}^{\mu }+k^{\mu }\mathcal{M}_{SD}^{\mu }=0$%
.

Let us now analyze the soft photon behaviour of this process. The form
factor is expanded around $k=0$ as 
\begin{eqnarray*}
\mathcal{F}\left( \left( p_{1}+k\right) ^{2},p_{2}^{2}\right) &=&\mathcal{F}%
\left( p_{1}^{2},p_{2}^{2}\right) +2p_{1}\cdot k\times \mathcal{F}^{\prime }+%
\mathcal{O}\left( k^{2}\right) \\
\mathcal{F}\left( p_{1}^{2},\left( p_{2}+k\right) ^{2}\right) &=&\mathcal{F}%
\left( p_{1}^{2},p_{2}^{2}\right) +2p_{2}\cdot k\times \mathcal{F}^{\prime }+%
\mathcal{O}\left( k^{2}\right)
\end{eqnarray*}
since we expect $\mathcal{F}^{\prime }=\frac{\partial \mathcal{F}}{\partial
p_{1}^{2}}\left( p_{1}^{2},p_{2}^{2}\right) =\frac{\partial \mathcal{F}}{%
\partial p_{2}^{2}}\left( p_{1}^{2},p_{2}^{2}\right) $ when evaluated
on-shell, i.e. at $p_{1}^{2}=p_{2}^{2}=m^{2}$. The amplitude is now 
\begin{equation}
\mathcal{M}_{IB+SD}^{\mu }=\mathcal{F}\left( p_{1}^{2},p_{2}^{2}\right)
\left[ \frac{p_{1}^{\mu }}{p_{1}\cdot k}-\frac{p_{2}^{\mu }}{p_{2}\cdot k}%
\right] +\mathcal{O}\left( k\right)  \label{LowStruct}
\end{equation}
i.e. the terms of order $\left( k\right) ^{0}$ disappeared. This result is
valid in full generality, it does not depend on the present structure
dependent term (\ref{SDTerm}) chosen for illustration (see \cite{Low}, \cite
{NonCstF}).

Now let us analyze the soft photon limit for the imaginary part (\ref{ImVD}%
). We have obtained for the two cuts the structures 
\begin{eqnarray*}
\mathcal{V}^{\mu } &=&\mathcal{F}\left( p_{1}^{2},p_{2}^{2}\right) \left[ 
\frac{A^{\mu }}{k}+B^{\mu }\right] \\
\mathcal{D}^{\mu } &=&\mathcal{F}\left( p_{1}^{2},p_{2}^{2}\right) \left[ 
\frac{C^{\mu }}{k}\right] +\mathcal{O}\left( k\right)
\end{eqnarray*}
Except for the $\mathcal{O}\left( k\right) $ terms in $\mathcal{D}^{\mu }$,
this is exactly the soft photon expansion one would get for a constant form
factor. Hence we expect from the analyzes of previous sections that the
combination will vanish when $k\rightarrow 0$. Note that it is essential
that no constant term (due to the momentum dependence of $\mathcal{F}$)
appears in $\mathcal{M}_{IB+SD}^{\mu }$. This was guaranteed by the gauge
invariance of $\mathcal{M}_{IB+SD}^{\mu }$, which in turn necessitates the
contribution of some structure dependent amplitude.

\end{document}